\documentstyle[12pt,aasms4]{article}
 \def\today{\number\day\enspace
      \ifcase\month\or January\or February\or March\or April\or May\or
      June\or July\or August\or September\or October\or
      November\or December\fi \enspace\number\year}
 \def\clock{\count0=\time \divide\count0 by 60
     \count1=\count0 \multiply\count1 by -60 \advance\count1 by \time
     \number\count0:\ifnum\count1<10{0\number\count1}\else\number\count1\fi}
 
 \def\newline{\hfil\break}
\def\et{{\it et~al. }}

 \def\bul{\ifmmode\bullet\else$\bullet$\fi}

 \def\deg{\ifmmode^\circ\else$^\circ$\fi}

 \def\kms{\ifmmode \hbox{ \rm km s}^{-1} \else{ km s$^{-1} $}\fi} 

\def\sec{\ifmmode \hbox{\rm sec}\else{sec}\fi} 
\def\yr{\ifmmode \hbox{\rm yr}\else{yr}\fi} 
\def\myr{\ifmmode \hbox{\rm Myr}\else{Myr}\fi} 
\def\gyr{\ifmmode \hbox{\rm Gyr}\else{Gyr}\fi} 

\def\hz{\ifmmode \hbox{\rm hz}\else{hz}\fi} 
 
\def\kpc{\ifmmode \hbox{\rm  kpc}\else{kpc}\fi} 
 \def\pc{\ifmmode \hbox{\rm  pc} \else{pc}\fi} 
\def\mpc{\ifmmode \hbox{\rm Mpc} \else{Mpc}\fi} 

\def\erg{\ifmmode \hbox{\rm erg} \else{erg}\fi} 

 \def\yrs{\ifmmode \hbox{\rm yrs}\else{yrs}\fi}

 \def\angstr{\ifmmode{\rm \AA}\else\AA\fi}

\def\ho{\ifmmode H_0\else$H_0$\fi}
\def\omo{\ifmmode\Omega_0\else$\Omega_0$\fi}
\def\to{\ifmmode T_0\else$T_0$\fi}
\def\h-1{h^{-1}}

\def\d2#1#2{{d^2#1 \over d#2^2}}

\begin{document}

\title{The Survival of The Core Fundamental Plane Against 
Galactic Mergers} 
\author{Kelly Holley-Bockelmann and Douglas Richstone \altaffilmark{1}}
\affil{Astronomy Department, University of Michigan}
\altaffiltext{1}{and Institute for Advanced Study, Princeton, NJ 08540}

\lefthead{Bockelmann \& Richstone}
\righthead{Survival of the Core Fundamental Plane}

\authoraddr{Ann Arbor, MI 48109-1090}
%\\
%\authoraddr{kelly@astro.lsa.umich.edu, dor@astro.lsa.umich.edu}

\begin{abstract}

The basic dimensional properties of the centers of elliptical
galaxies, such as length scale, luminosity, and velocity dispersion,
lie on a Fundamental Plane similar to that of
elliptical galaxies as a whole.  The orientation of this plane, and the
distribution of core parameters within it, project to a strong
correlation of core density with either core or total luminosity, and
indicate that low luminosity ellipticals are much denser than high
luminosity galaxies (HST data suggests that this relationship may be as
steep as ${\rho_c} \propto L^{-2}$).  In addition, low luminosity ellipticals
have a much smaller length scale than their high luminosity counterparts.
Since we think small
galaxies are occasionally accreted by big ones, the high
density of these galaxies and their likely durability against the
time-varying tidal field of the bigger ones suggests that they will
survive substantially intact in the cores of larger galaxies and would
be easily visible.  Their presence would destroy the observed
correlation.

Motivated by this apparent inconsistency between an observed fact and
a simple physical argument, we have developed an effective
simulation method and applied it to the problem of the accretion of very
dense secondary companions by tenuous primaries.  We have studied the
accretion of objects of varying luminosity ratios, with sizes and
densities drawn from the Fundamental Plane under the assumption that
the mass distribution in the central parts of the galaxies follows the
light. The results indicate that in mergers with mass ratios greater
than 10, chosen with an appropriate central density dependence on 
luminosity, the smaller object is only stripped down to the highest density 
encountered in the primary during the accretion process.  
Thus, the form of the core Fundamental Plane suggests that
the mass distribution in the galaxy centers is different than the
light distribution, or that an understanding of secondary survival 
requires more than the dynamics of visible stars.

\end{abstract}

\section{Introduction}

In the framework of hierarchical galaxy formation,  
massive elliptical galaxy centers are 
merger archives, in the sense that the history of each component is
imprinted upon it.  

The stellar Roche limit suggests that the accretion of dense galaxies
by larger, more tenuous ones would leave dense galaxies nearly
intact. Hence, the central density of any massive galaxy should vary
widely as a function of the extent of prior merging. Instead, the
global parameters of elliptical galaxies vary simply with the
luminosity.  Elliptical galaxies exist in a well-defined slice of
global parameter space called the Fundamental Plane (Dressler \et 1987, Djorgovski and Davis 1987). Similarly, the {\it central}
parameters of galaxies are also arranged on a plane called the core
Fundamental Plane (Faber 1997).
 
On the core Fundamental Plane, small galaxies can be $10^6$ times more
dense than large galaxies (Faber, 1997). This enormous dynamic range
in density presents a formidable challenge for hierarchical merging
because in this theory, the dense center that results from the 
nearly intact accretion of a small galaxy by a bigger one will cause 
the remnant to
evolve off the Fundamental Plane.  Consequently, the persistence
of the Fundamental Plane is an indication either that mergers are
uncommon or that additional forces are at work to destroy the
secondary.  

There is considerable evidence that mergers do occur.  Observations of
brightest cluster galaxies indicate that massive ellipticals evolve by
accretion of small satellites at a rate of up to $0.2 L_{\star}$ per
Gyr (Lauer 1988, Blakesee and Tonry 1992). Simulations of galaxy
cluster evolution also demonstrate that mergers occur, although the
importance of this process in the creation of cD galaxies has been
debated (Malumuth and Richstone, 1983; Merritt, 1984).  So, while
it is clear from the persistence of the Fundamental Plane that mergers
must destroy the smaller satellites, it is not clear how such dense
systems are destroyed.

Recently, Weinberg (1997) has argued that internal heating of the
secondary can cause it to evaporate well inside the tidal radius.
This heating occurs when some of the orbits of stars in the secondary
resonate with the time-dependent tidal forces imposed by the primary.
To demonstrate this, Weinberg perturbs a solution of the Boltzmann
equation and recomputes the gravitational field of the secondary for
King models on the global Fundamental Plane. He states that a
satellite with a mass ratio as high as 100:1 will dissipate entirely
by the time it falls to the primary center (Weinberg, 1997). As we
will show later, although we agree with Weinberg's result for the
specific problem he did, his models lack the density contrast observed
for Fundamental Plane galaxy cores, so his simulations have limited
application to the problem of core Fundamental Plane persistence.

Mergers of galaxies with very different densities are expensive to
simulate with a straightforward N-body approach due to the disparity 
in two important timescales.  First, we must consider the orbital decay
of the secondary into the primary, which is of order of the dynamical
friction timescale, and varies as $ M_1 / M_2 \ \rho_1 ^{-1/2}$. 
We must also
consider the orbital periods of stars within the secondary, which vary 
as $\rho_2 ^{-1/2}$ where $\rho_1$ and $\rho_2$ are the densities of
the primary and secondary. So, even for a modest mass ratio, tracking 
the stars in each
galaxy over the entire merger requires
on the the order of $100 \  M_1/M_2 \ (\rho_1 /\rho_2)^{-1/2}$ 
timesteps.  There are few published direct
simulations of this type of merger for this reason.  In one important
case, however, Balcells and Quinn (1990) did simulate an elliptical
galaxy merger with a mass ratio of 10:1 (and density ratio of
approximately 1:14).  Contrary to Weinberg's result, the secondary
remained intact inside its tidal radius as it sank to the center of
the primary, although the orbital decay was perhaps too fast to 
produce resonant heating.

We have designed an approach to determine the conditions under which
core Fundamental Plane secondaries survive mergers. This approach can
efficiently merge systems with mass ratios as high as 1000, a range
unexplored by standard simulations. In this paper, we describe this
method and apply it to the problem of secondary survival. We review
the core Fundamental Plane in $\S$ 2, our galaxy models in $\S$ 3.1,
the approximation method in $\S$ 3.3, and our code in $\S$ 3.4.  The
tests of the method and results of our simulations can be found in $\S$ 4.
Briefly, it appears that core Fundamental Plane secondaries survive
mergers in which the mass ratio is 10:1 or larger.  Section 5
discusses the implications of our results on the persistence of the
core Fundamental Plane and previews future work.

\section{The Core Fundamental Plane}

In the parameter space composed of the effective radius, $r_e$, the
effective surface brightness, $\mu_e$, and the central velocity
dispersion, $\sigma_0$, elliptical galaxies are confined to a plane
--- the ``Fundamental Plane'' or FP (Dressler \et (1987), Djorgovski
\& Davis, 1987).  Recently Jorgensen \et (1996) characterized the FP
as $\log r_e = 1.24 \log \sigma_0 - 0.82 I_e + \gamma$, where $\gamma$
may depend on environment.  A popular interpretation of the
Fundamental Plane is that it is a manifestation of virial equilibrium
together with an orderly and slow dependence of galaxy mass--to--light
ratio on luminosity.  Faber \et (1997) report the existence of an
analogous plane for elliptical galaxy centers based on core
parameters, though this core Fundamental Plane has more scatter than the
global Fundamental Plane.  The existence of such a plane indicates
that the processes which act upon galaxies are primarily
scale-independent.  More importantly, this plane provides an important
constraint on the merger history of elliptical galaxies, because if a
galaxy merges, the remnant must also lie upon the plane.  In
particular, this implies that when a small dense galaxy collides with
a massive galaxy, the remnant must have the low density center that is
characteristic of a massive primary.

When simulating the mergers of real galaxies, then, it is imperative
that the initial models lie on both planes. Although the Fundamental
plane is 2-dimensional, we adopt the convention of Faber (1997) and
represent the core galaxies as a 1-dimensional family dependent on
total luminosity.  This is a convenient approximation which is made
possible by the tight correlation between global and break luminosity
and by the nearly edge-on projection of the plane onto these axes.
The core data can be fit as follows:

\begin{eqnarray} \log \ r_b & = & - 0.46 \, M_v - 7.7 \\
 \log \ L_b & = & 1.3 \, \log L_e - 5.097 \\
 \log\ r_{\rm eff} & = & 0.75 \, \log \ r_b + 2.06,  
\end{eqnarray}
where $r_b$ 
is the break radius, $M_v$ is the absolute visual magnitude, $L_b$ is
the visual luminosity at the break radius, $L_e$ is the visual
luminosity at the effective radius, and $r_{eff}$ is the effective
radius.  The following additional relations were obtained (Faber \et
 1997) by assuming a slightly varying mass-to-light ratio 
along the Fundamental Plane of $M/L \propto {L_e ^{0.25}}$,  
\begin{eqnarray}
  \sigma_0 & \propto & L_e ^ {0.2} \\
 \rho_b & \propto & \sigma_0^2 /r_b^2  \propto L_e^{-1.9}.  
\end{eqnarray}
This last equation is the crux of the difference between our
simulations and Weinberg's. Weinberg's expression of the Fundamental
plane is $\rho_{b} \propto M^{0.5}$, where M is the mass of the
galaxy, so for reasonable choices of ${M}/{L}$, his Fundamental Plane
has a much shallower dependence on density with radius. See figure 1
for best-fit projections of the core Fundamental Plane to the Lauer
\et  (1995) core data. 

\section{Methods}
\subsection{Modeling the Core Fundamental Plane}

\def\rhotilde{\stackrel{\sim}{\rho}}

The disruption of the secondary galaxy centers will depend
on the density ratio between the primary and the secondary, and on their density profiles.  We
therefore construct models designed to reflect the range in profiles
observed on the core Fundamental Plane.  We use $\eta$ models for this 
purpose  (Tremaine \et 1994, Dehnen, 1994).  These are simple, 
spherical systems defined by their density distributions: 
\begin{equation} 
\rho_\eta (r) = {M_{tot} \over {r_b}^3} \rhotilde (\eta, r/r_b), 
\end{equation}
where 
\begin{equation} 
\rhotilde (\eta, x) 
  	\equiv {1\over {4 \pi}}
 {\eta \over {x^{3-\eta} (1 + x)^{1+\eta}}}, 
\qquad 0 <  \eta \le 3,  
\end{equation}
where $\eta$ is the single dimensionless parameter that defines the
slope of the inner density profile, $\rho$ is the stellar density,
$M_{tot}$ is the total mass, $r_b$ is the break radius, and x is the dimensionless radius $x =
r/r_b$.

We fit the the deprojected stellar density profiles provided by
Gebhardt \et (1996) to determine $\eta$ as a function of the absolute 
magnitude $M_v$.  The data (see Figure 1) show a correlation of the 
structural parameter $\eta$ with $M_v$. The ridgeline of the 
correlation was fit by eye as follows: 
\begin{equation} 
\eta = \cases 
{ 3.0 - 0.375 M_v - 9.1, &if $M_v  \le -19.5$;\cr
  3.0 - 0.15 M_v - 4.7, &if  $M_v \> -19.5$}. 
\end{equation}

While eta models are well-suited to cover the range of luminous matter
densities in elliptical galaxy centers, the model's density profile
drops off more rapidly ($\rho \propto r^{-4}$) than a real galaxy's at
large radii. Although we anticipate that the envelope of a secondary
would be completely stripped in an encounter with a primary, and is
therefore irrelevant, we were concerned that truncating the envelope
might have an unforseen effect on our calculations. So, we mimicked an
$r^{1/4}$ law outer profile by constructing a basic set of galaxies with
eta model envelopes. We call these galaxies double $\eta$
models. Their density is defined as follows: 
\begin{equation} 
\rho(r) = {{M_c} \over { {r_c}^3}} \, \rhotilde(\eta_c, r/r_c) + {{M_e} \over {{r_e}^3}} \, \rhotilde(\eta _e, r/r_e), 
\end{equation}

\noindent where the subscript c and e corresponds to eta model
parameters chosen at the core and effective radius, respectively. The core
radius in equation 9 is chosen to be mimic the 
break radius on the 
core Fundamental Plane. The envelope term parameters are 
chosen to be consistent with
global Fundamental Plane parameters and with $ r_e \propto {r_c}^{1.33}$
(see equation 3) (Faber, 1997). For $r_{e}$ on the order of $100 \ r_c$,
$\rho(r)$ provides a reasonable approximation to a deVaucoleur profile.

\subsection {Populating the Galaxies}

\def\ee{\varepsilon}

To populate our galaxies, we require the phase space distribution function. 
From our spherically symmetric initial density profile, we can
derive the initial distribution function, $f(\ee)$, 
via Eddington's formula:
\begin{equation} 
f(\ee) = 
{1\over {\sqrt 8}{\pi^2}} 
\left[ 
\int_0^{\ee}
{d^2\rho \over d \Psi^2} 
\ {d\Psi \over  \sqrt{(\ee - \Psi)}} 
+ 
{1\over \sqrt\ee}
\left({d\rho\over d\Psi} \right)_{\Psi = 0} 
\right], 
\end{equation}
where $\ee = {-E + \Phi_0}$ is the relative energy, and ${\Psi} =
{-\Phi + \Phi_0}$ is the relative potential (Binney and Tremaine,
1987). $\Phi_0$ is chosen so that the distribution function is positive
for $\ee > 0$ and zero for $\ee \le 0$. The second term in equation
12 is zero, because $\Psi = 0$ when $ r \rightarrow \infty$, and as $
r \rightarrow \infty$, $\rho \propto r^{-4} \propto\Psi^4$. Determining 
$f(\ee)$, then, simply requires knowledge of
${{d^2\rho} / {d \Psi^2}}$.

\noindent
Defining $u = 1/r$, 

\begin{equation} 
 {d^2 \rho \over d \Psi^2}  
	= {d \over du} \left( {{d\rho}\over {d\Psi}} \right) 
		\bigg/ {d\Psi \over du}.
\end{equation}

\noindent
For unscaled $\eta$ models, 

\begin{equation}
{d \over du}{\left[ d\rho \over d \Psi \right]}
	= { \left[
{12 u^2 + 4(3-\eta) u^3 \over z} - 
{ 4 u^3 + (3-\eta)u^4 \, z^\prime 
\over {z^2}} \right] }, 
\end{equation}
\noindent
where ${z} = {(1+u)^2}$, and ${z^\prime} = {{dz} / {du}}$.
\begin{equation} 
{d \Psi \over du} =  
\cases{ 
(1 + u)^{-\eta}, & \mbox{\rm if} $\eta \ne 1.0,$ \cr
 1/(1+u),       & \mbox{\rm if} $\eta = 1.0$}.
\end{equation}

\noindent
See Tremaine et al, 1994 for an equivalent expression of the
second derivative. The second derivative for the double $\eta$ model case
has many more terms, but follows directly from the above derivation. 
Given the distribution function, we know the number of particles within
the phase space  $d^{3}{\rm{ v}} \ d^{3}{\rm{x}}$. To populate the 
secondary, then, we
choose a random scalar radius from the mass profile, and determine the
speed by choosing a random scalar velocity from:

\begin{equation} 
f(v)dv= 4\pi v^2 f(\Psi - {1\over 2 } v^2) dv. 
\end{equation}
Finally, cartesian position and velocity components are 
derived from uncorrelated random orientations of {\bf r} and {\bf v}. 

These two component galaxies were simulated with 5000 particles.  In
these simulations, the outer envelope was stripped away on the first
encounter with the center of the primary, and we took advantage of this
result by conducting additional simulations of the inner portions 
of the galaxies with an inner
eta model of 2000 particles. This markedly improved our spatial
resolution. These simulations contain a reasonable particle number,
but in principle, the simulations can and will be repeated with more 
particles as computational speed increases. Model parameters are 
given in Table 1.  Figure 1
illustrates the placement of the models on the core Fundamental Plane.

\subsection{The Tidal Force on the Secondary}

A straightforward technique to simulate mergers is to follow the
interaction of the primary and secondary with an N-body code. However,
when the galaxies involved have very different masses, the interaction
takes an unreasonably long time to simulate in this straightforward
manner, as described in the introduction.  We have developed an
approximation method which dramatically decreases the computational
time required to simulate high mass contrast mergers.  Our
approximation hinges on the assumption that the potential of a high
mass galaxy is not significantly altered by the impact of a low mass
object. This approximation allows us to concentrate on the behavior of
the secondary only. The merger, then, can be characterized as the
secondary's response to the external force it experiences as its 
orbit decays in the field of the primary. 

Clearly, the external force on the secondary depends upon its
position in the primary galaxy. We determine the orbital decay trajectory
by solving the equations of motion for a point particle in the field 
of a galaxy of mass $M_1$: 

\begin{equation} \vec  F_{\rm 1-2}(\vec S) = - {{{G M_1(\vec S) m_2} \over { {\vec S}^2}}} - {F_{\rm {fric}}(\vec S)}, \end{equation}

\noindent
where $\vec S$ is the vector from the center of the primary and to the secondary center, and
$ F_{\rm {fric}}$ is a modified Chandrasekhar dynamical friction
formula. This also presupposes
an unchanging primary, and neglects the effect of mass loss in the
secondary.

\begin{equation} F_{\rm {fric}}(\vec S) = - f_{\rm drag} \space {{{4 \pi {\rm ln} \Lambda G^2 \rho_{1} {m_2}^2} \over { \vec {v_{2}}^3}} \Bigl[ {\rm {erf}}(X) - {{{2X}\over{\sqrt{\pi}}}e^{-{X}^2}}\Bigr] \space \vec v_{2}}, \end{equation}

\noindent
where $\vec v_{2}$ is the velocity of the secondary's center of mass, 
$X \equiv v_{2} / ( \sqrt{2 \sigma})$, $\sigma = \sqrt {0.4 G M_{1}/r_{1, {\rm eff}}}$, $\Lambda$ is the Coulomb logarithm which was set to $M_{1}/M_{2}$, and
$f_{\rm drag}$ is a drag coefficient. This drag coefficient is necessary,
 because
the decay times generated by  
numerical treecode experiments were longer and exhibited more pericenter
passes than the unmodified Chandrasekhar formula. Since the number 
of pericenter
passes is correlated with the amount of destruction in a particular 
secondary, we
chose to match the number of pericenter passes and the overall decay time
by adjusting the analytical dynamical friction force by
drag coefficient $f_{\rm drag}$. Figure 3
shows the treecode-generated trajectory and several analytically-derived
trajectories with different drag coefficients.

Once the orbital decay trajectory is computed, the external 
force on particles within the secondary can be written as:

\begin{equation} \vec F_{\rm S}(\vec r) = - {{G M_1(\vec S+ \vec r) m_2} \over {(|\vec S + \vec r| + \epsilon)^3}}, \end{equation}

\noindent
where $\vec r$
is the vector which points from the secondary center to a secondary particle, and $\epsilon$ is a softening parameter which we chose to be close to 
the core radius of the
secondary. We advance the system in the secondary frame, so the appropriate 
force is the sum of the tidal force and the secondary's self-gravity:

\begin{equation} \vec F_{tot}(\vec r) = \vec F_{in}(\vec r) + \vec F_{\rm S}(\vec r) -\vec  F_{\rm S}(0). \end{equation}

We tested the assumption that the primary is unchanged by the secondary
for each mass ratio by 
dropping a secondary point mass into a 10000 particle primary using
either Hernquist's treecode (1987) or self-consistent field code (1992).  
The density profile of the primary did not change for the 100:1 and 10:1
cases (see Figure 2), so the approximation is a valid one for those
mass ratios. This method is
more computationally efficient as the mass ratio increases, and is over 50
times 
faster than the Hernquist treecode in the 10:1 mass ratio.

\subsection{Approximating the Self-Gravity of the Secondary}

We constructed a self-consistent particle-field code which treats the
mass distribution as a sphere and uses a logarithmic grid. This force is estimated
by first convolving the mass distribution with an adaptive kernel function
(Silverman, 1986) to get a density estimate. An adaptive kernel
provides a non-parametric density estimate on a grid. We use an
Epanechnikov kernel (an inverted parabola) for the density 
estimate:
\begin{equation} 
K_e = \cases{
{3\over 4( 1 - r^2)} & \mbox{\rm if $r < 1$ }, \cr 
           0         & \mbox{\rm otherwise},}  
\end{equation}
where $r= |x-x_i|/h$, x is a particle position, $x_i$ is a grid point, and h is the
window width. An initial density estimate is obtained, then the window
width of each grid element is adjusted according to the initial
density such that low density regions have a large window width, and
high density regions have greater mass resolution. This gives us a
smoothed $M(r)$ from which force and potential are easily computed (Merritt, 1996).  Although
it is true that this code, which forces spherical symmetry, is not an
appropriate choice if one wishes to model the shape of the highly
non-spherical process of satellite disruption, we were concerned about
whether a particle was bound to the secondary, not with obtaining an
accurate shape for the stripped debris. Furthermore, since the tidal
force is a very strong function of r, the core of the secondary, where
the calculation matters most, is spherically symmetric.

Particle trajectories are advanced in the secondary frame with a
single timestep leapfrog method. 

We tested this code by following the total
energy of each secondary without the influence of an external force.
For double $\eta$ secondaries, energy was conserved to within  1 $\%$ 
over 1700 core crossing times, the time in which the 100:1 mass
ratio case would have decayed had an external force been present. The mean
energy and root mean square energy of the individual stars were conserved
to within 2 $\%$ over this interval.
Inner $\eta$ models exhibited 
more heating, initially, because the kernel inaccurately determined the 
density very near the center of the system. Since inner $\eta$ models 
have a better resolution, there were a small number of particles 
that were close enough to the center of 
the secondary for this inaccuracy to matter. We restored energy
conservation by setting the mass distribution to that of the inner
$\eta$ model inside $0.01 r_b$, and reflecting any particles
entering this sphere back outward with the same energy and
angular momentum. These modifications to the code affected 
approximately 10 particles, but markedly improved the energy efficiency.
See figure 4 for $\delta \space E_{\rm tot}$ vs. crossing time.

\subsection {Initial Conditions} 

Each secondary was followed as an isolated system for several 
dynamical times to ensure its virial equilibrium. 
The merger simulation begins
as the secondary is launched from $3 r_{\rm half} $ of the primary
with an initial velocity that depends upon the orbit desired; for 
a plunging orbit,
we launch the secondary with zero velocity. 
We define an angular momentum parameter ${\kappa} \equiv {{L} / {L_{\rm circ}}}$,
where L is the initial orbital angular momentum, and $L_{\rm circ}$ is
the orbital angular momentum that the secondary would have
if it were on a circular orbit about the primary at the launch point.
Each simulation 
is followed until the secondary's apocenter is less than $0.1 \ r_b$ of the
secondary.  

\section{Results}

Since our goal is to investigate the breakup or survival of dense 
secondaries orbiting in primaries, we initially focused on the 
encounters we thought would be the most destructive.  The destruction of 
the secondary, however, must occur without 
leaving a lasting imprint
on the remnant, and this depends on where in the
primary a secondary is 
stripped. For example, in a decaying 
circular orbit, the tidal force from the primary would gain 
strength as the secondary's orbit decays.  In this
case, even if substantial damage occurred, the debris would continue 
to orbit near the center of the primary, and the density of the remnant would be large near the center. The orbits most effective for dispersing the secondary over a large volume must be the deeply plunging ones, since these encounters would carry the secondary past the 
center of the primary on its first orbit with
considerable orbital energy.  Our basic set of 
experiments, therefore, were 
parabolic encounters with no angular momentum. For an overview of
the different experiments conducted, see table 2.  

\subsection{Basic Set of Experiments}
Anticipating that the most damaging effect would be due to tidal
forces experienced as the secondary passes through the center of the
primary, we launched a series of nearly parabolic, plunging orbits.
In this basic set of experiments, we investigated three mass ratios:
100:1, 10:1, and 2.5:1. These mass ratios correspond to density
ratios of approximately 1:51, 1:20, and 1:1 at a radius of 1 pc, 
and the density ratio increases at smaller radii.
The secondary was modeled as a double
eta model in order to represent the envelope. We focus first on the 
100:1 simulation. Figure 5 shows the time evolution of the secondary
in the force field of the primary.  Here, we learn that most of the
mass is stripped at pericenter, and comes off impulsively in a cloud
which continues to expand as the secondary crosses the primary center
again. In figure 6, we illustrate the change in the secondary center
from beginning to end.  The secondary remained intact inside the 
tidal
radius, $ {r_{\rm tide}}^ 3 \approx  {{M_{1}} / { M_{2}}} \ {S}^3 $.
Figure 7 illustrates the change in the secondary for the 10:1
mass ratio.  In the 2.5:1 experiment, the secondary and primary have
similar central densities, and the secondary is destroyed (figure 8).

We note that the primary and secondary in the 2.5:1 mass ratio are
close enough in mass for the approximation scheme to fail, because
the primary is not expected to remain unaffected by a merger of similar
mass. We address this fact by conducting a treecode experiment,
which we discuss in $\S$ 4.5.

\subsection{Comparison of Drag Coefficients}

Recall the drag coefficient, $f_{\rm drag}$ in $\S$ 3.3 which
adjusts the strength of Chandrasekhar's dynamical friction force
to be comparable to treecode generated orbital decay times.
Changing the strength of the frictional force, though, also
changes the velocity of the secondary, and therefore both the
magnitude of the shock on the secondary as it passes through pericenter and the number of pericenter passes. 
We tested the importance of this
effect by simply varying the drag coefficient on the 100:1
fiducial experiment. In 2 simulations with 
$f_{\rm drag}= 0.1$ 
and $f_{\rm drag}=0.2$,
we saw no significant difference in the secondary's inner density 
profile, although it is apparent that the secondary suffers more tidal 
stripping as it goes through more pericenter passes (see figure 9).
Consequently, we set $f_{\rm drag}$ to the value determined
by a treecode simulation of a point secondary in a responsive primary for each mass ratio in all further experiments.
This provided a more accurate timescale for the decay, but did not
affect the survival of the secondary.

\subsection{Comparison between Inner $\eta$ and Double $\eta$ Models}

To achieve better sampling of the central regions of the secondary,
we populated only the center of the galaxy with an inner $\eta$ model.
While this method excludes the secondary envelope, it
is a reasonable plan because our basic set of experiments 
determined that the envelope is stripped away quickly without
markedly changing the density profile at the center
(see $\S$ 4.1 and figures 6-8). Note that the orbital decay due
to dynamical friction is the same as the previous experiment. 
We only followed the stars in the inner part of the galaxy but used the total mass of the secondary to determine the orbital decay.

As in the fiducial set of experiments, these simulations show
that the secondary remains intact in the 100:1 and 10:1 mass ratio,
and that the secondary breaks up in the 2.5:1 mass ratio (figures 10-12).

\subsection{Non-radial Secondary Orbits}

To explore the effect that different
orbits have upon the disruption of the secondary, we launched
secondaries in the 10:1 mass ratios on
${\kappa} \equiv {{L} / {L_{\rm circ}}}
= 0.2 $ and $\kappa = 0.6$ orbits, and the 100:1 
mass ratio on a $\kappa = 0.2$ orbit. The ratio of second apocenter to 
first pericenter is approximately 66 in the 100:1, $\kappa = 0.2 $ orbit,
and 34 and 16 for the 10:1, $\kappa = 0.2$ and $\kappa = 0.6$ 
orbits, respectively. The computation
time for these simulations required us to use inner $\eta$ models
and to follow only the bound particles. Still, the $\kappa = 0.6$
run took approximately 2.5 months. In every case, the secondary's
innermost density profile remained essentially intact; the 
secondary center was steeper and more dense than the primary, so the
center of the remnant is dominated by the secondary (Figures 13-15).
For a comparison of the effect that orbit has upon the disruption of 
the secondary, we overplot the final secondary profile for each orbit
in the 100:1 and 10:1 inner $\eta$ experiments (figure 16).

\subsection{Treecode Comparison}

Our method works best at high mass contrasts, because for a given
primary, the lower the secondary mass, the less the primary is
affected by a merger.  In addition, the gravitational force on the
secondary is better approximated as a tide when it is small compared
to the self-gravity of the secondary, as it is for a small, dense,
secondary.  On the other hand, the treecode is superior at low
mass-contrasts because it can follow changes in the primary. 
We compare the results from these two problems at a mass ratio of  2.5:1
We launched a 2000 particle double $\eta$ secondary into a
5000 particle double $\eta$ primary along the same orbit as described
in the fiducial experiments and followed the behavior with the
Hernquist treecode.

In the treecode simulation the primary center is highly disrupted by
the secondary. Our method cannot reproduce this effect. In spite of the marked difference
between the behavior of the primary in each method, the qualitative
behavior of the secondary is the same: the secondary central density
decreases to that of the primary (figure 17). The fact that we obtain
the same qualitative results provides some confidence in the method. In addition, the fact that our method fails
to address the change in the potential of the primary in a clear
warning that the method is quantitatively unreliable when the
mass ratios are very close to 1.

\subsection{Comparison to Weinberg's Results}

A key test of our method is to duplicate Weinberg's experiment using
our approximation method. Weinberg chose to represent the Fundamental
Plane with King model galaxies (King, 1966). 

King models are a single parameter set of solutions of the Vlasov 
equation, which can be labeled by the dimensionless potential 
${W_0} = - V_0 /(2\sigma^2)$ (where $V_0$ and $\sigma$ are 
the central potential and dispersion parameter), or equivalently 
by the ratio of ``tidal'' to core radii $r_t/r_c$.  The mass 
distribution in a King model is given by $M(r) = \mu (r/r_c) \rho_0 
r_c^3$, where $\rho_0$ is the central density and $\mu$ is a 
dimensionless function.  Weinberg adopted a single choice 
of $W_0$ in his experiments, so his models all had the same 
dimensionless mass distribution but scaled up or down in both 
density and length scale.

We chose to duplicate one of Weinberg's 10:1 mass ratio simulations.
In this experiment, he populated two $W_0 = 9.5$ King models with
densities defined by the global Fundamental Plane: $\rho \propto M^{ -
{0.5} }$. He determined the orbital decay with
Chandrasekhar's analytical dynamical friction formula, using the
Coulomb logarithm ${\ln \Lambda} = {max
\lbrack{ \ln({R_{\rm circ}}/{r_{\rm half}}),0.1}\rbrack}$, which
caused the orbit to decay in 10 orbits for a choice of $\kappa =
0.1$. We prefer our handling of the orbital dynamical friction because
it is calibrated by the treecode. However, we wish to critically
compare our calculation of the effect of the fluctuating tide to
Weinberg's problem, so we selected a choice of $f_{\rm drag}$ such
that our orbit also decayed in 10 passes for $\kappa = 0.1$. In all
dimensionless characteristics, our experimental set up is the
same. See table 3 for the parameters used in this experiment.
\footnote{ There is a typographical error in the
Weinberg preprint that causes in an inconsistency between the King
model concentration and the tidal radius; this prevented us from
taking Weinberg's precise radius in parsecs and his Coulomb
logarithm. We circumvented this problem by creating two $W_0 = 9.5$
King models that differed in mass by 10 and density by a factor of 3, 
consistent with Weinberg's model.}

Weinberg contends that the secondary is disrupted by the time it
reaches the center, with most of the disruption due to stripping near
the primary center. In this particular case, we obtain the same result
(figure 18); that is, the secondary is stripped by nearly a factor of
10, and $\rho_2 = \rho_1$.
This shows is that when we do exactly the same experiment as Weinberg,
we get the same result. Therefore, the differences between our 
results and Weinberg's are not due to differences in calculation methods.
Instead, we believe the differences are due to different choices of 
galaxy models and the use of these models to mimic the Fundamental Plane.
In our paper, we carefully selected each galaxy to fall on the global
and core Fundamental Planes, assuming light follows mass within the
galaxies.

\section{Discussion}
\vskip0.5cm

We have developed an approximation scheme that allows us to
investigate the mergers of systems with high density ratios.  We can use
this method to simulate merging systems with central density contrasts as high
as 1500 with less than a month of computer time on an UltraSparc.

Our simulations show that disruption will not occur for secondaries
that are much denser than their primaries.  While resonant heating may
enhance tidal shocking in the disruption of secondary galaxies,
this only appears to matter when the density contrast between primary
and secondary is not too large.  Weinberg's King model secondaries
evaporated because they do not possess the high densities that real
secondaries exhibit on the core Fundamental Plane.

This result presents a quandary. We suspect, from observations of
merger rates, that dense secondaries do merge with more diffuse
primaries. In the process, the secondary must be destroyed, since
observations of the centers of massive merger remnants rarely show
dense, small secondaries. Instead, nearly every large remnant has a
low-density center with a density profile that flattens out inside a
break radius; this is a key aspect of the core Fundamental Plane. 
The inability of purely stellar mergers to destroy the secondaries 
suggests that an important component of the problem is missing.

One possible solution to this quandary may be massive black
holes. The widespread presence of massive black holes in galaxy 
centers has been suggested, for example, by Kormendy \&
Richstone (1995), Magorrian (in press), 
Richstone (in press).   
Black holes act as excellent scatterers, and their presence in
the centers of massive primaries may provide tidal forces sufficient
to disrupt a dense secondary. We will present extensive simulations of the
effect of black holes in a second paper. However, alternative possibilities 
include radial
variations of M/L within individual galaxies, with an amplitude
strongly dependent on galaxy luminosity, in the sense that
luminous galaxies possess sharp concentrations of unseen mass. In order
to be effective, the mass concentration would have to be more
dramatic than the M/L variation with luminosity implied by
the core Fundamental Plane. 

Support for this work was provided by 
the Space Science Telescope Institute, through general observer grant 
GO-06099.05-94A, and by NASA through a theory grant G-NAG5-2758.
DR thanks the Guggenheim Foundation and the Ambrose
Monell Foundation for support at the IAS.  We thank the members of 
the NUKER collaboration for helpful conversations. We also thank
the referee, Martin Weinberg, for his insightful suggestions.

\newpage

\figcaption[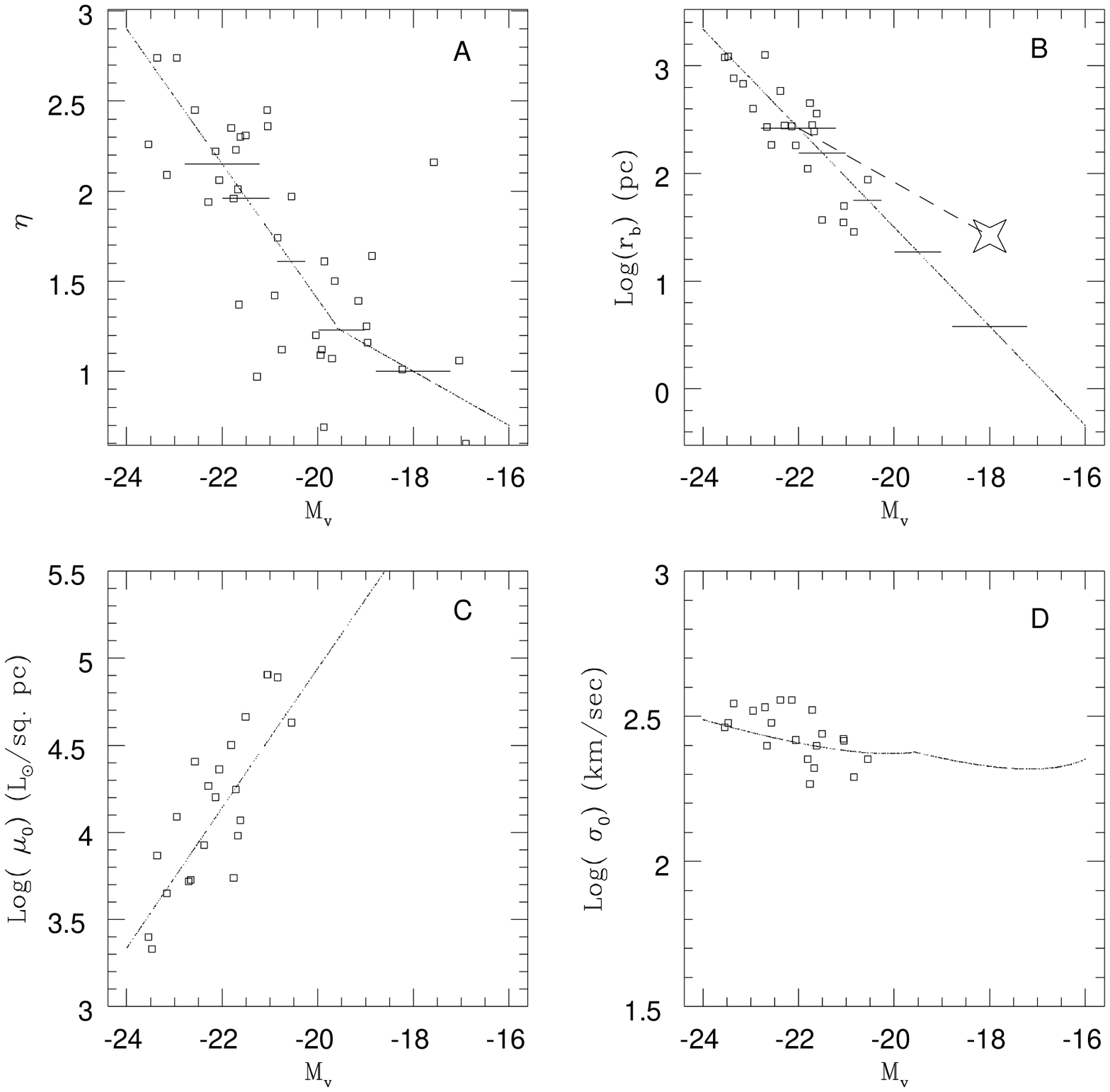]{A comparison of data and our model of core 
parameters of galaxies. In each panel, the Lauer \et (1995) 
galaxy core data are plotted as squares, and a fit of
the core and the power-law data is plotted as a line. Panel A shows estimated values of $\eta$ for the Lauer \et sample, together with our adopted model.
The 5 X's in panel A correspond to the 5 unique galaxies in Table 1.
The other three panels
represent the projections of the core Fundamental Plane for these
galaxies. Panel B is a plot of core radius versus absolute 
visual magnitude for the Lauer et al data and our model. In this panel, 
the stellated X demonstrates 
the core radius that
Weinberg's 1:100 mass ratio secondary would have if the primary were
to lie on the core Fundamental Plane at $M_v = -22$. 
This illustrates that Weinberg's simulated 
Fundamental Plane has a shallower slope, as seen by the dashed line.
Panel C plots resolution-limited surface brightness versus
absolute visual magnitude of the Lauer et al sample, with the central 
surface brightness and absolute visual magnitude of our models.
Panel D plots the resolution limited velocity dispersion versus
absolute visual magnitude for the data, and the central velocity dispersion
versus absolute visual magnitude of our models. 
\label{Figure 1}}

\figcaption[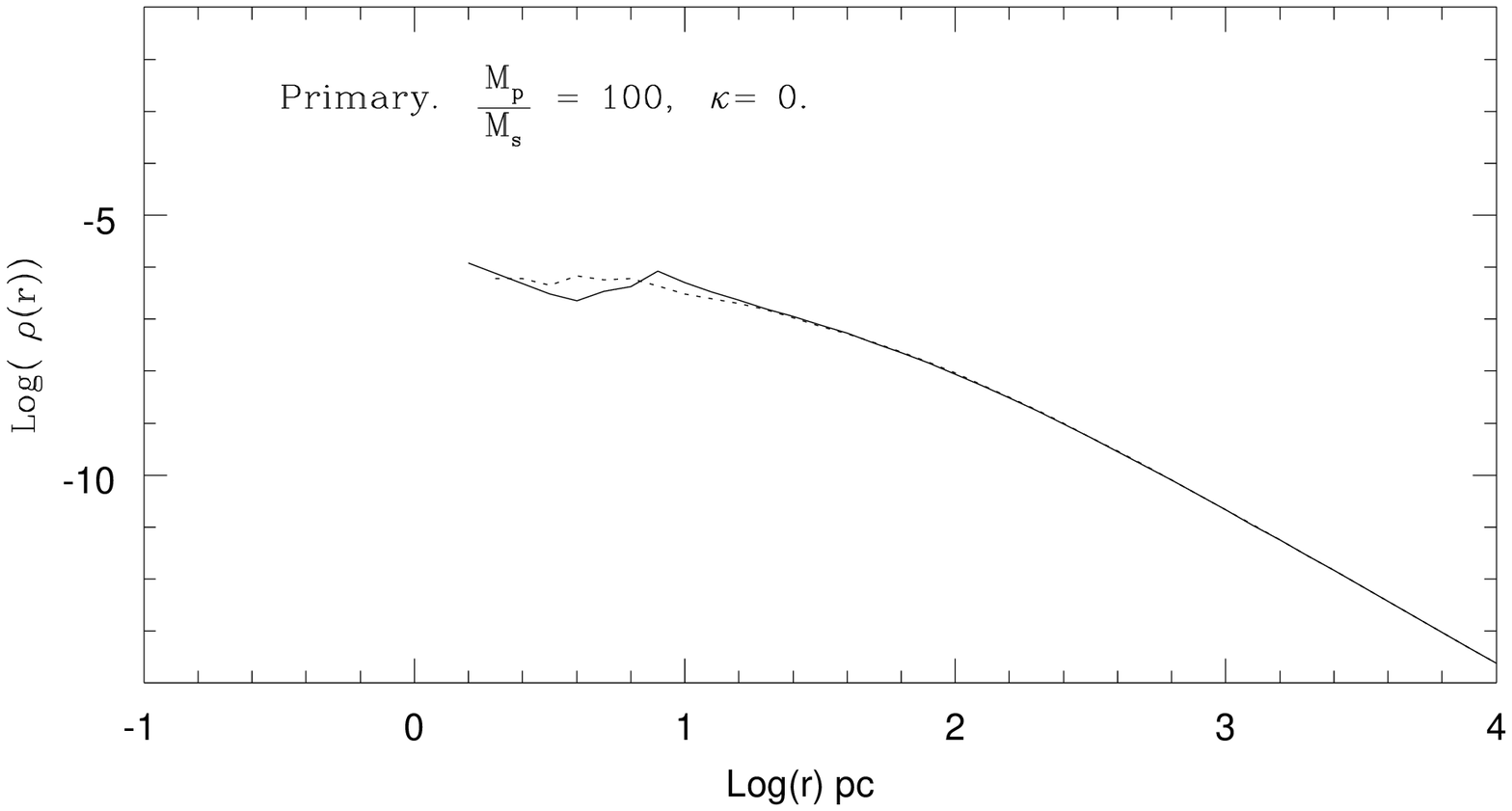]{The change in the primary during a 1:100 merger. The density of the primary is shown as a function of radius before and after the merger. The initial density of the primary is a dashed line, and the final density is a solid line. The changes are fluctuations due to small number
statistics. 
\label{Figure 2}}

\figcaption[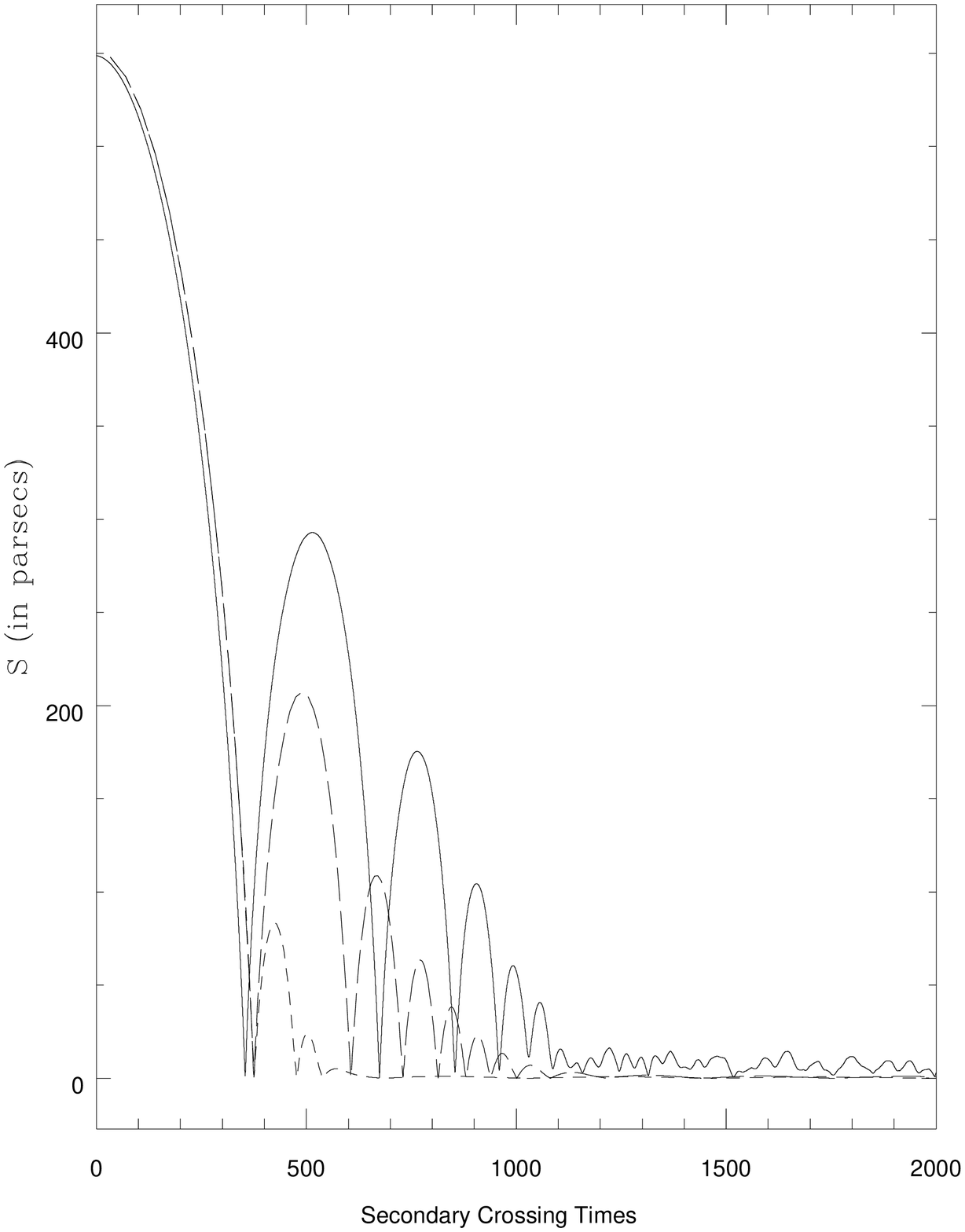]{The distance between primary and
secondary centers versus secondary crossing time. We plot the secondary
decay trajectory from a treecode experiment as a solid line, and the analytical dynamical friction derived decay trajectories from two experiments with 
varying drag coefficients. The short dashed line represents 
$f_{\rm drag} = 1.0$, and the long dashed line has $f_{\rm drag} = 0.2$.
\label{Figure 3}}

\figcaption[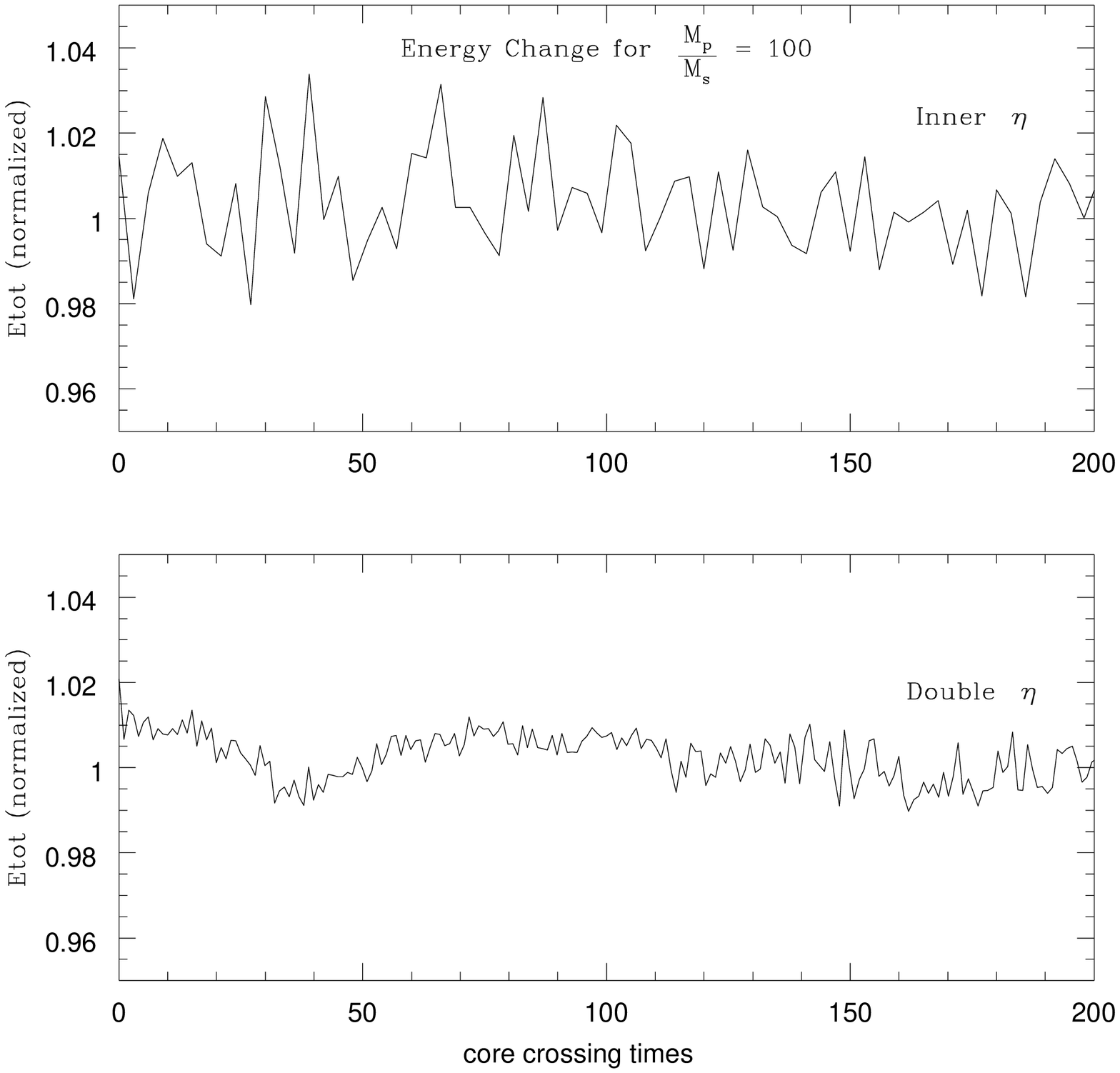]{Test of the method: energy conservation. 
We plot the total energy (normalized so that the mean energy is 1.0)
versus the core crossing time of the secondary for the inner
$\eta$ model secondary, and for the double $\eta$ model secondary.
Energy is constant to within 2 percent for both simulations, although
energy conservation is better for the double $\eta$ secondary after
it relaxes. \label{Figure 4}}

\figcaption[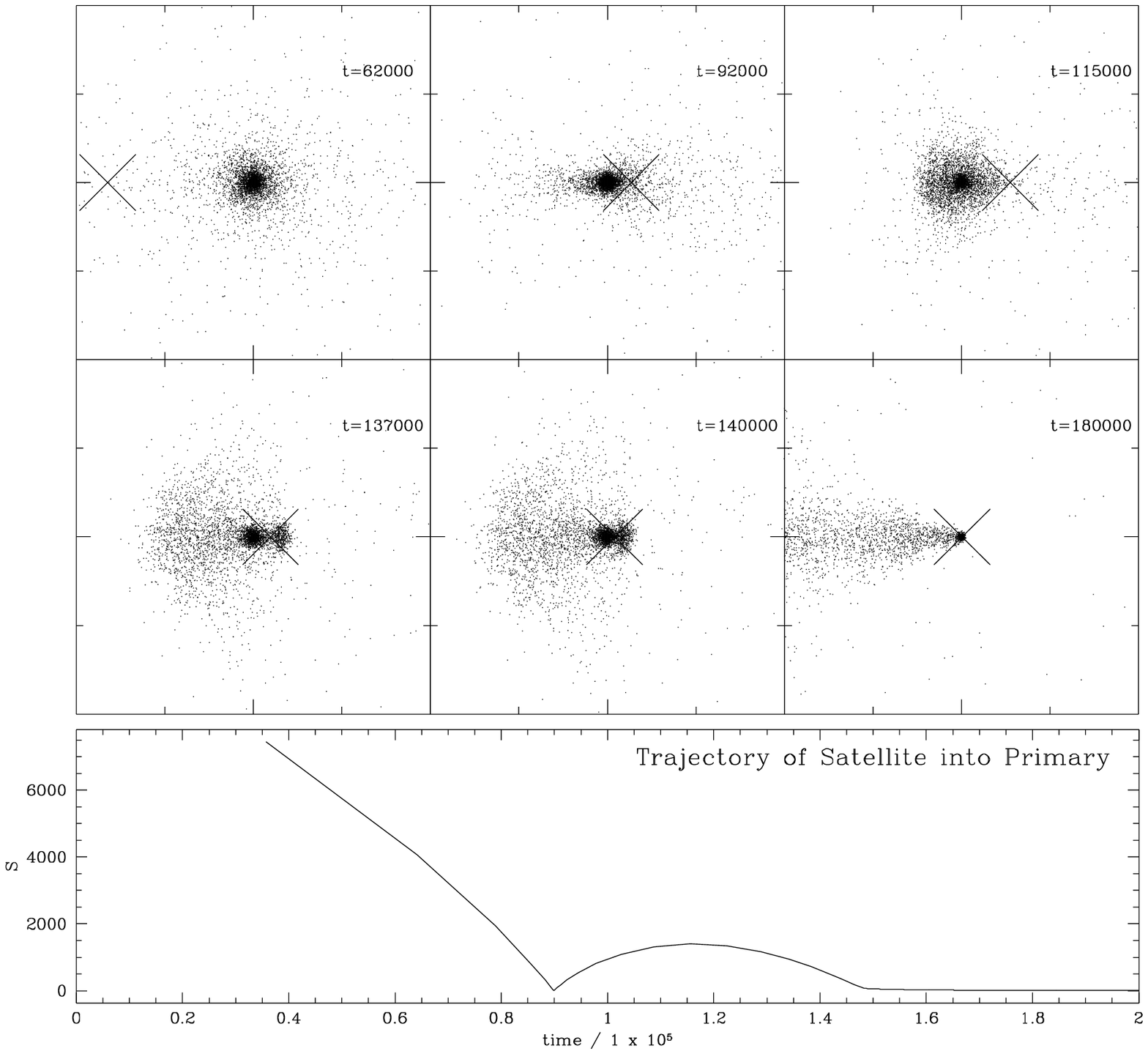]{The projection 
of the xy plane of a secondary as it merges with a primary 100
times more massive. Each panel represents a different snapshot of the
secondary along its orbital decay trajectory. 
The trajectory of the secondary is shown on the bottom of the plot, as in Figure 3. Most of the envelope particles
are unbound after the first pass, and by the second pass, $90 \%$
of the matter is stripped away. However, the inner particles
remain bound to the secondary's potential. \label{Figure 5}} 

\figcaption[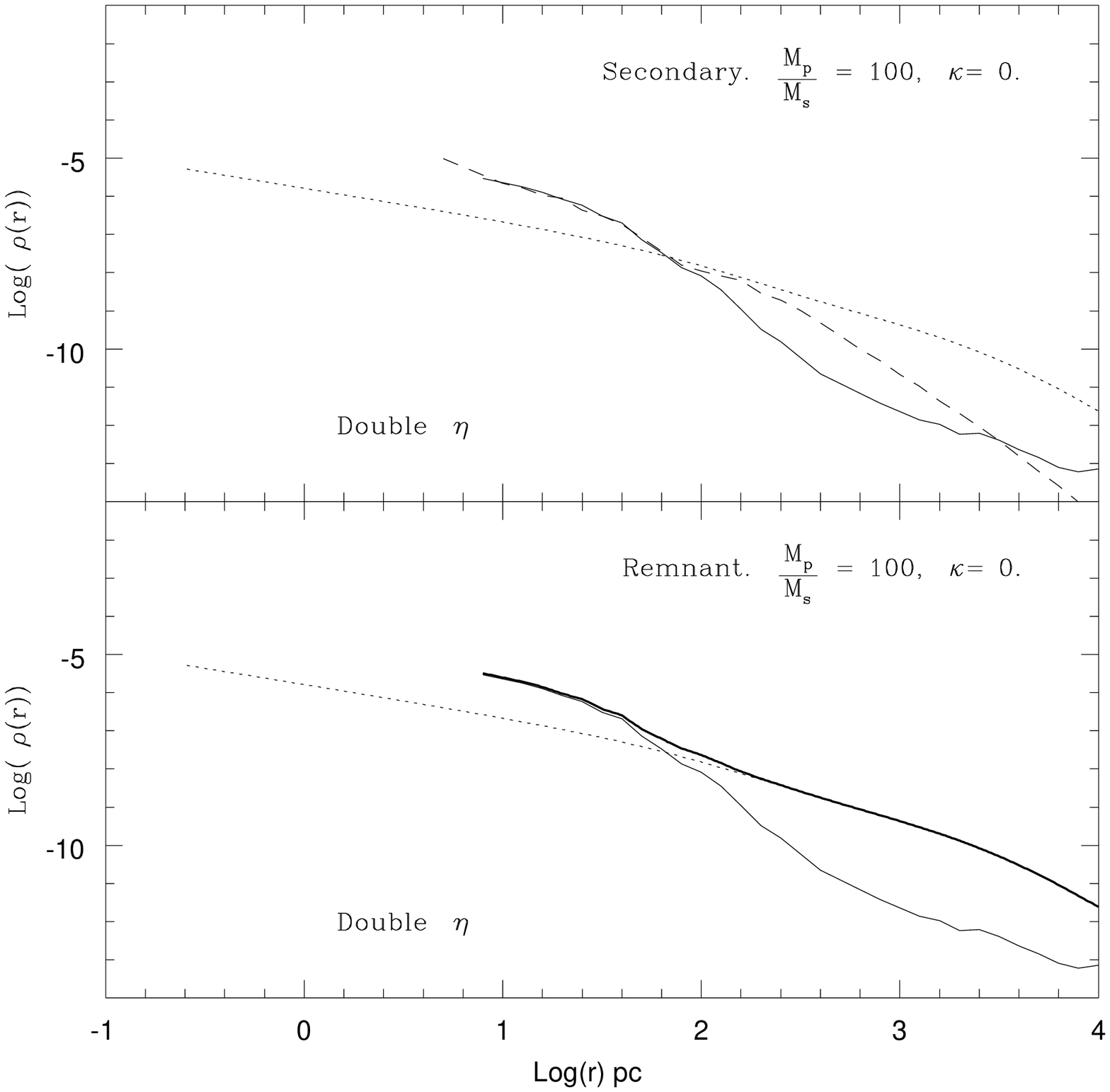] {Density profile for the mass ratio 100:1 basic experiment.
In the top panel, 
we illustrate the change in the inner density profile of the 
secondary. We plot the density at a radius r against radius in parsecs.
The solid line is the final secondary profile, the thick dashed
line is the initial state of the secondary, and the dotted line is
the density profile of the primary for comparison. In the bottom panel,
the thick solid line represents the resulting remnant, the dotted line corresponds to the final state of the primary, and the thin solid line
represents the final state of the secondary.
Inside the tidal radius, the secondary is unchanged (refer to $\S$ 4.1).
\label{Figure 6}}

\figcaption[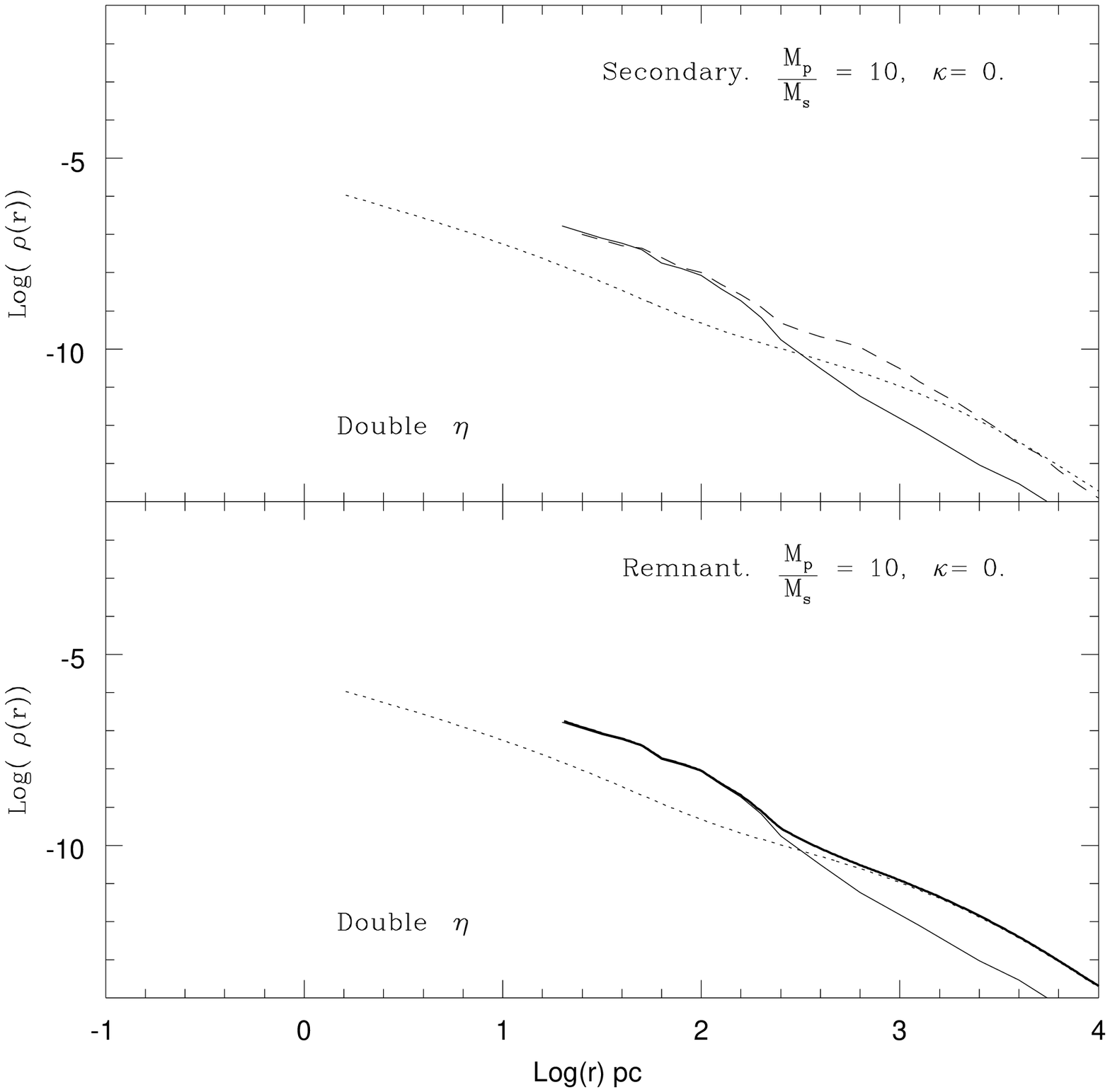] {Density profile for the 10:1 basic experiment.
See caption for Figure 6.
\label{Figure 7}}

\figcaption[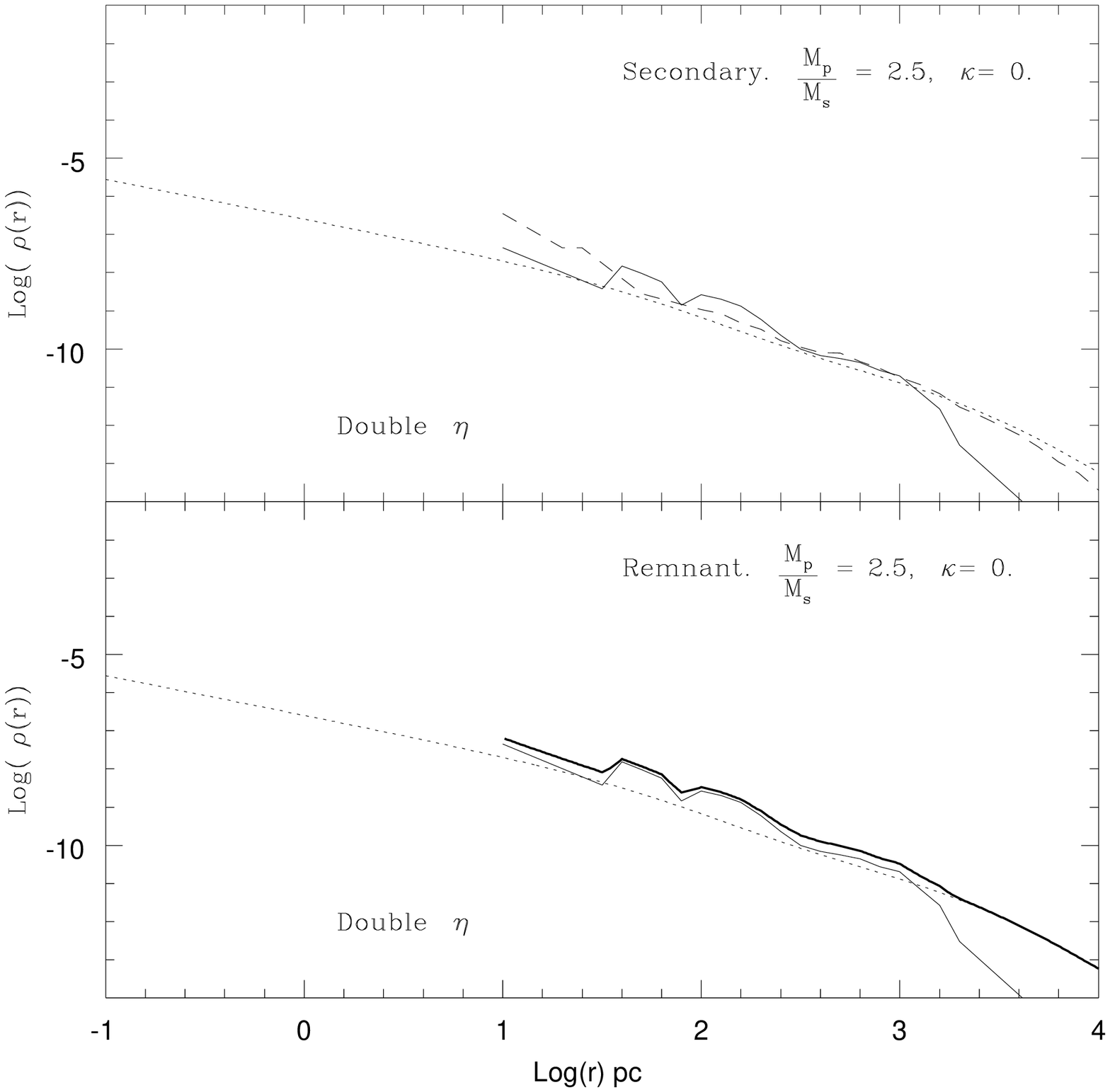] {Density profile for the  2.5:1 basic experiment.
See caption for Figure 6.
\label{Figure 8}}

\figcaption[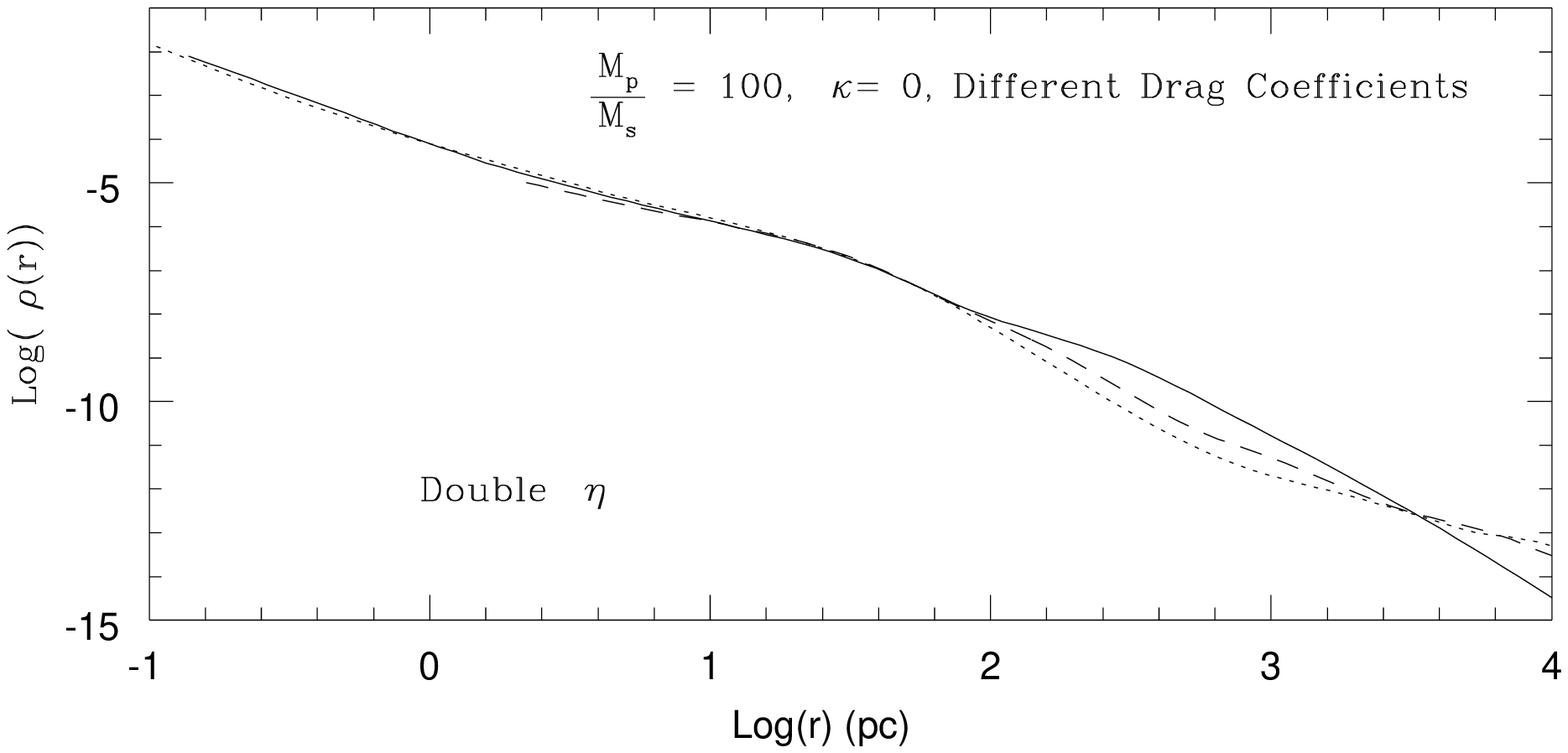]{We plot the secondary's density at radius r versus
radius for two drag coefficients.
We adjusted the magnitude of the 
dynamical friction force by a drag coefficient that we obtained 
by fitting the orbital decay time generated by Chandrasekhar to that
generated by a full treecode simulation.
This plot indicates that varying the drag coefficient, 
$f_{\rm drag}$, by a small amount does 
not affect the
overall result of the merger. The solid line is the initial secondary,
the dashed line is the $f_{\rm drag}=0.2$ case and the dotted line is for
$f_{\rm drag}=0.1$.
In both simulations, the
secondary remains intact at the center, although the more pericenter passes,
the more the outer envelope is stripped.  \label{Figure 9}}

\figcaption[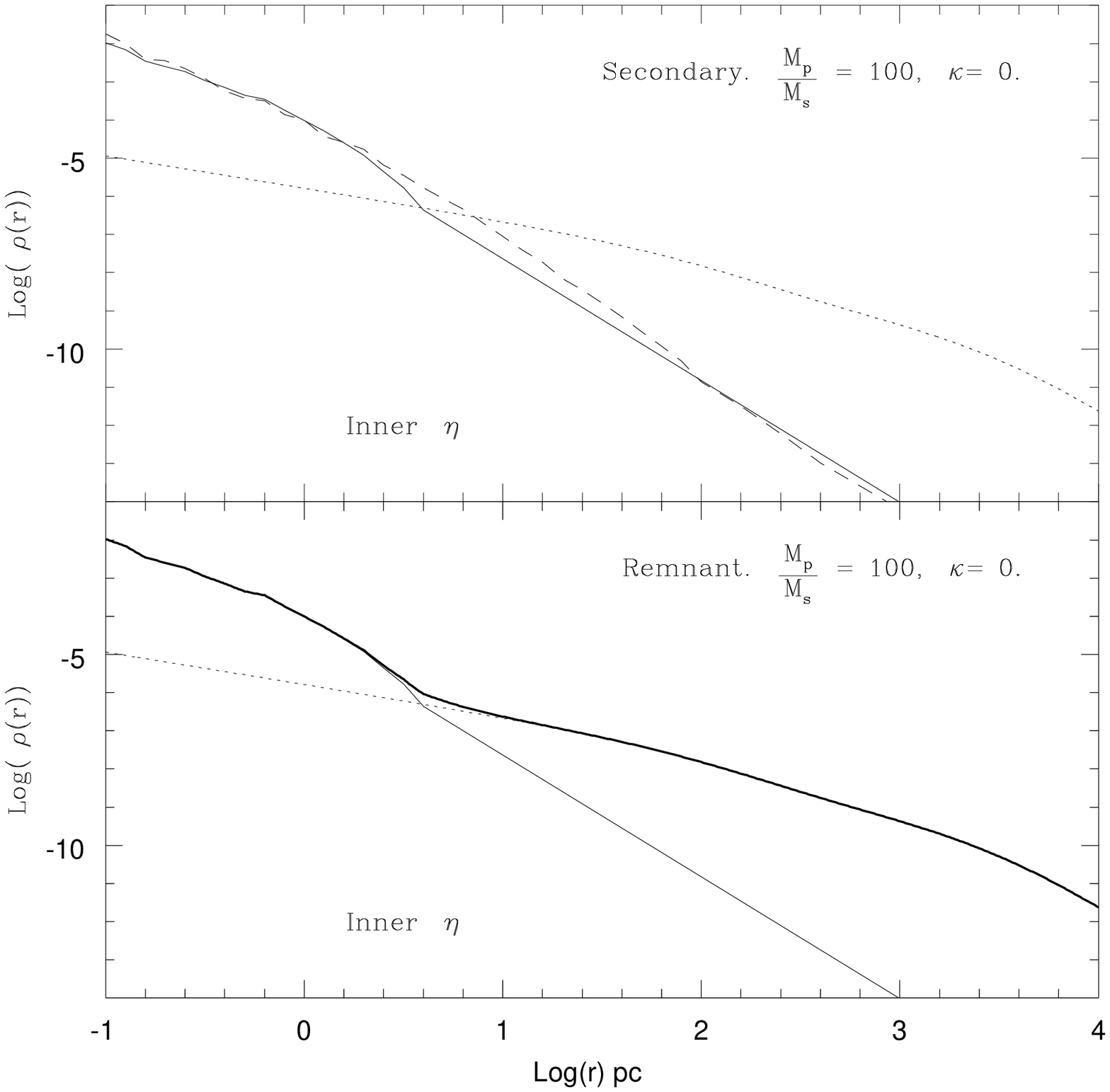]{Better inner resolution for 100:1.
See the caption for Figure 6.
Inside the tidal radius, the secondary is essentially intact, although a
comparison with Figure 6 demonstrates that more mass is lost inside 
the tidal radius (refer to $\S$ 4.3).
\label{Figure 10}}

\figcaption[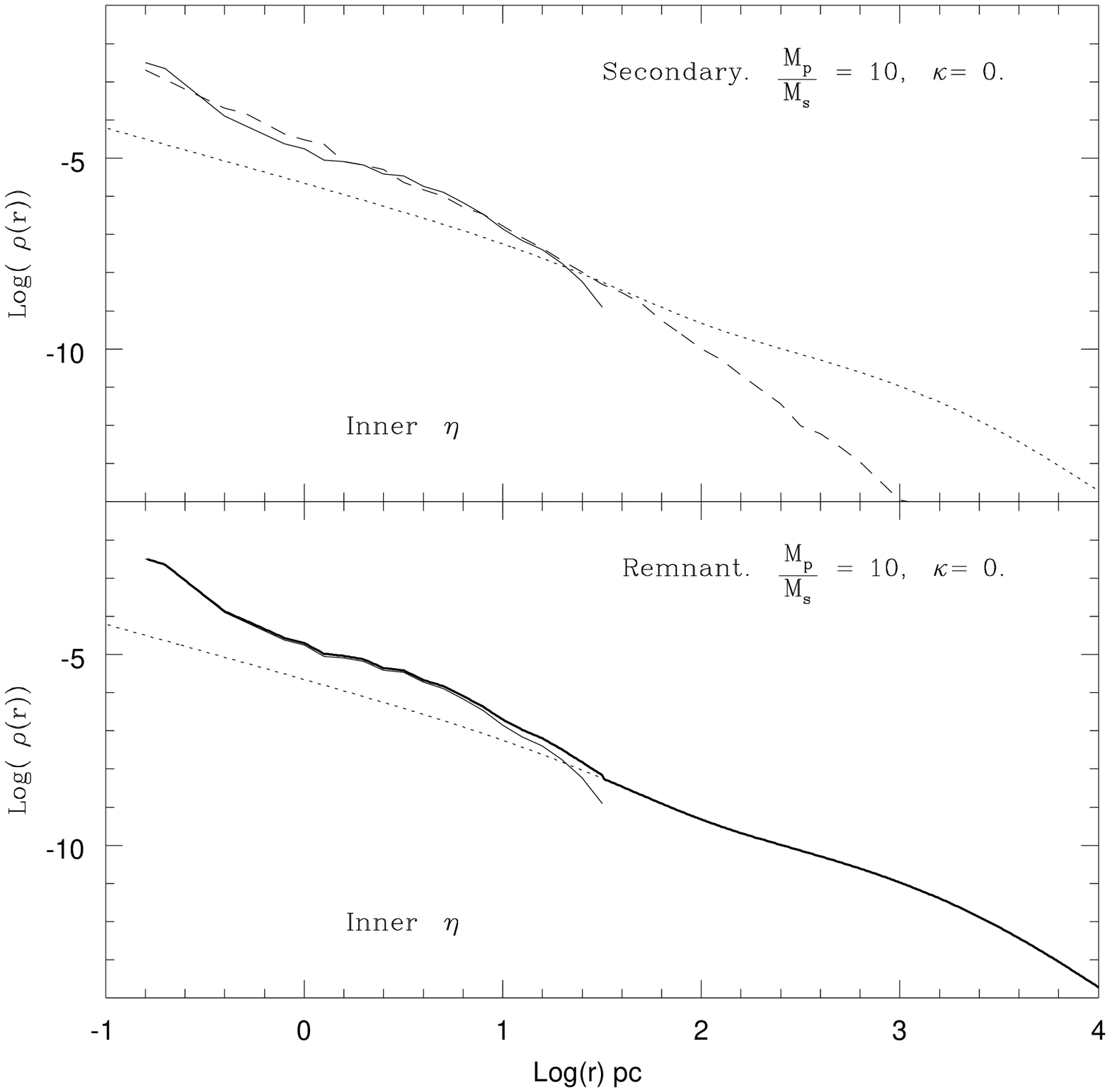]{Better inner resolution for 10:1.
See caption for Figure 6.
\label{Figure 11}}

\figcaption[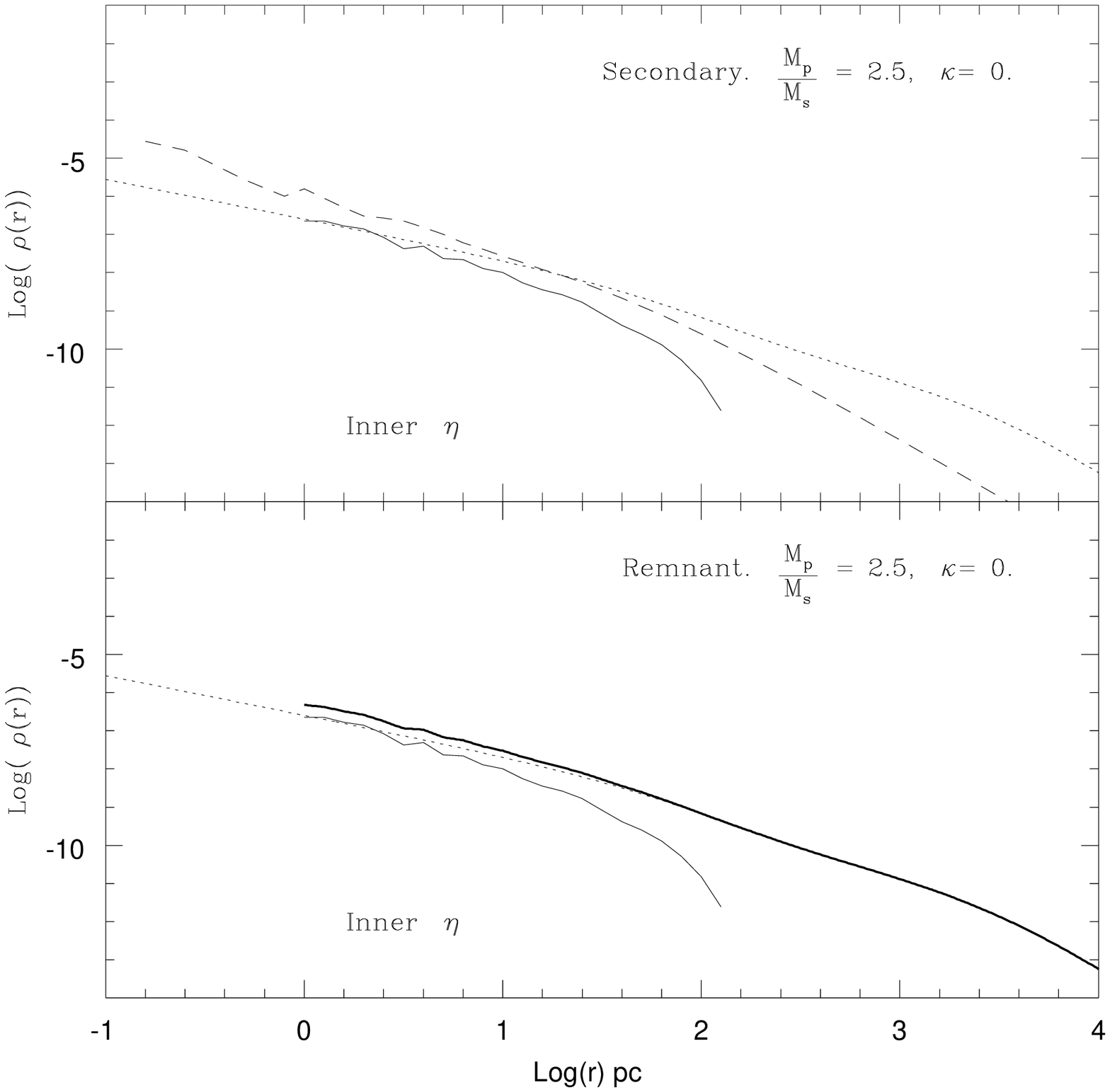]{Better inner resolution for 2.5:1.
See caption for Figure 6.
In this experiment, the density of the secondary at the center has dropped 
to or below the density of the primary.
\label{Figure 12}}

\figcaption[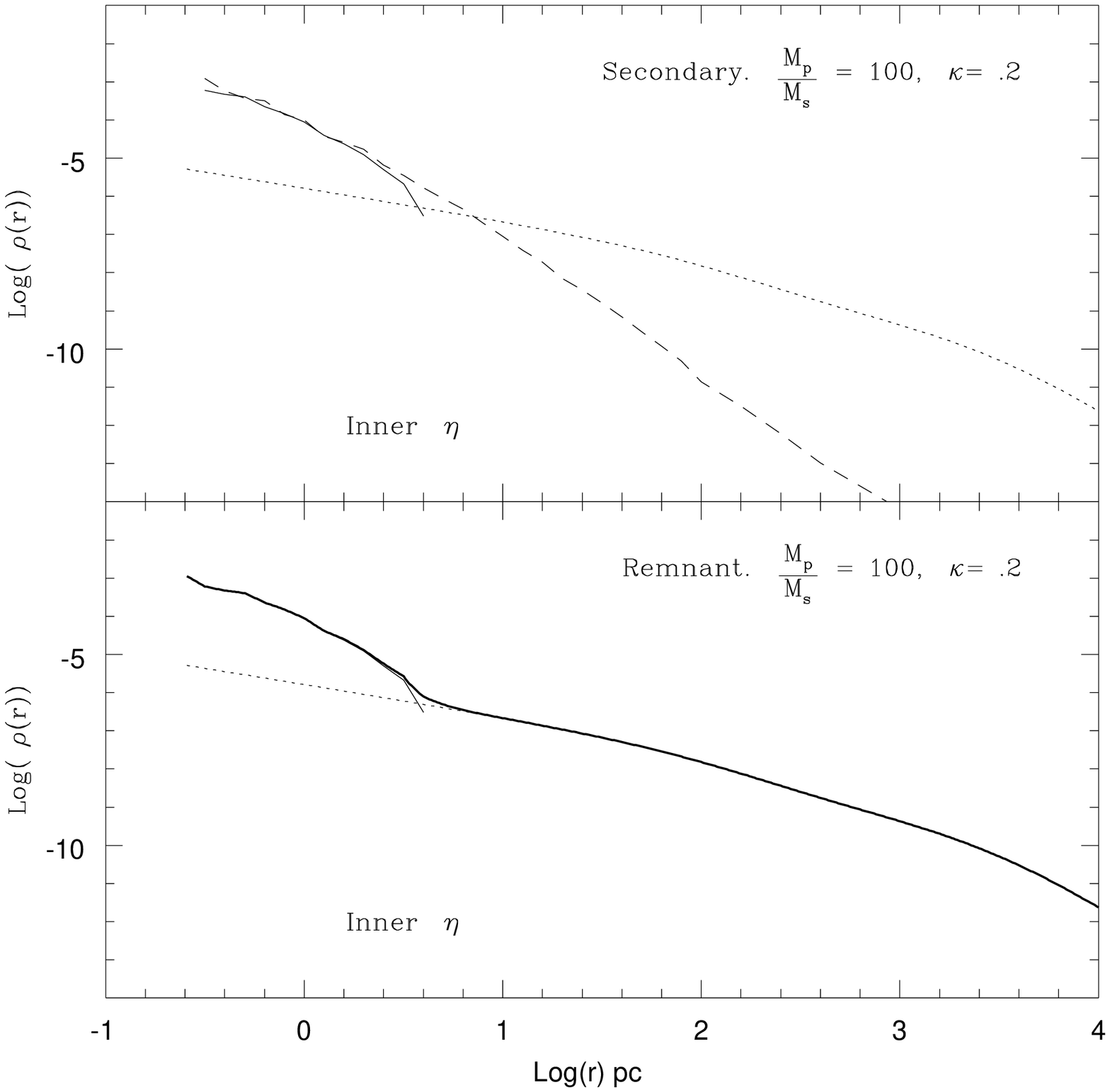]{$\kappa= 0.2$ orbit for 100:1.
See caption for Figure 6.
The secondary is intact inside the tidal radius.
\label{Figure 13}}

\figcaption[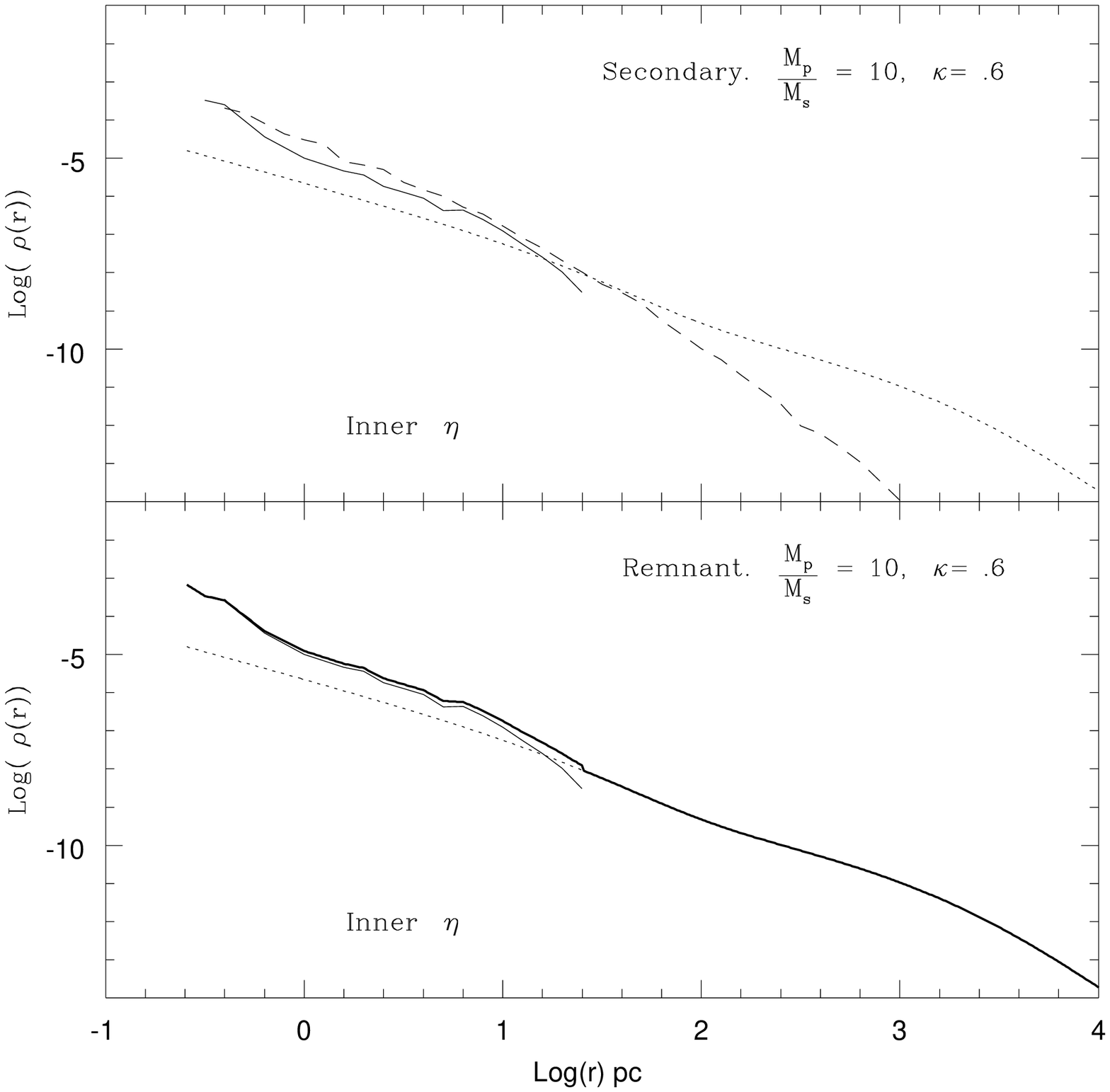]{$\kappa= 0.6$ Orbit for 10:1.
See caption for Figure 6.
In this experiment, there is a minor amount of mass loss inside the tidal 
radius, even for 
very small radii, but qualitatively, the
secondary survives and overwhelms the inner density profile of the remnant
(bottom panel).
\label{Figure 14}}

\figcaption[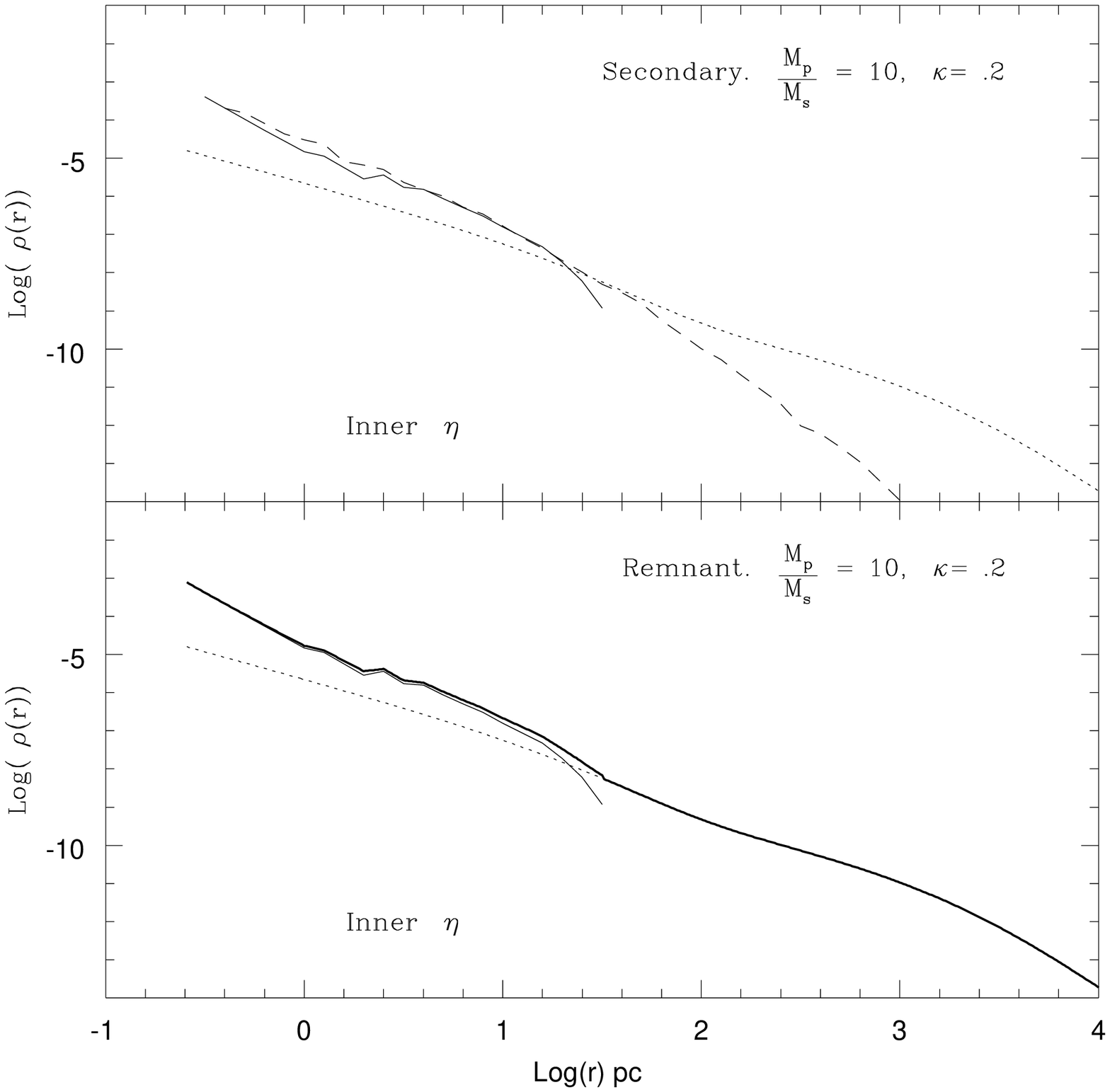]{$\kappa= 0.2$ orbit for 10:1.
See caption for Figure 6.
\label{Figure 15}}

\figcaption[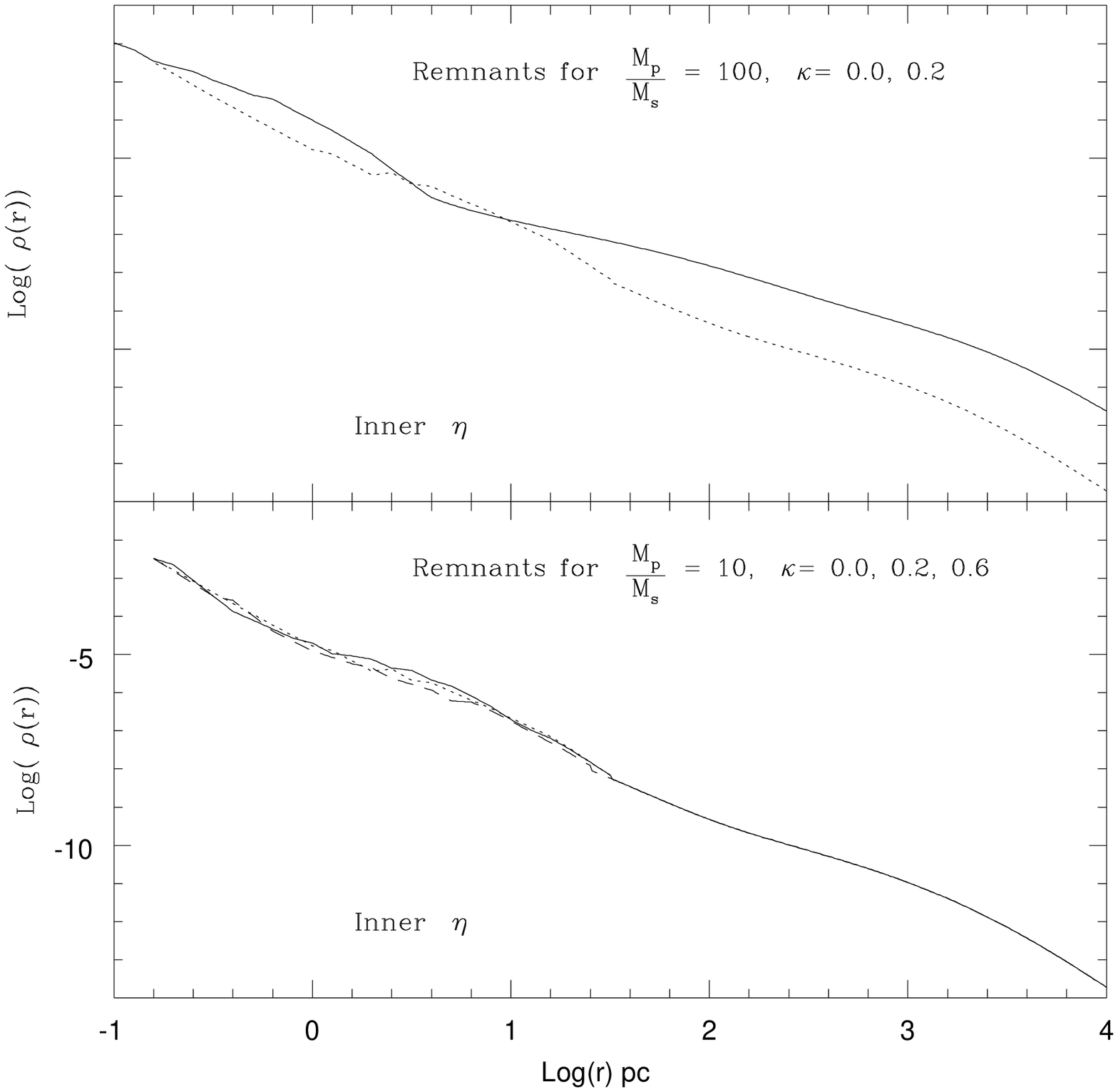]{Overplot of final secondary density for
different orbits. We plot the density at a radius r against radius in 
parsecs. The top panel is for the 100:1 mass ratio and the 
bottom panel illustrates the 10:1 mass ratio. The solid line in each
case is the $ \kappa=0.0 $ case, the short dashed line is the $ \kappa =0.2 $
case, and the long dashed line is the $\kappa =0.6 $ case. \label{Figure 16}}

\figcaption[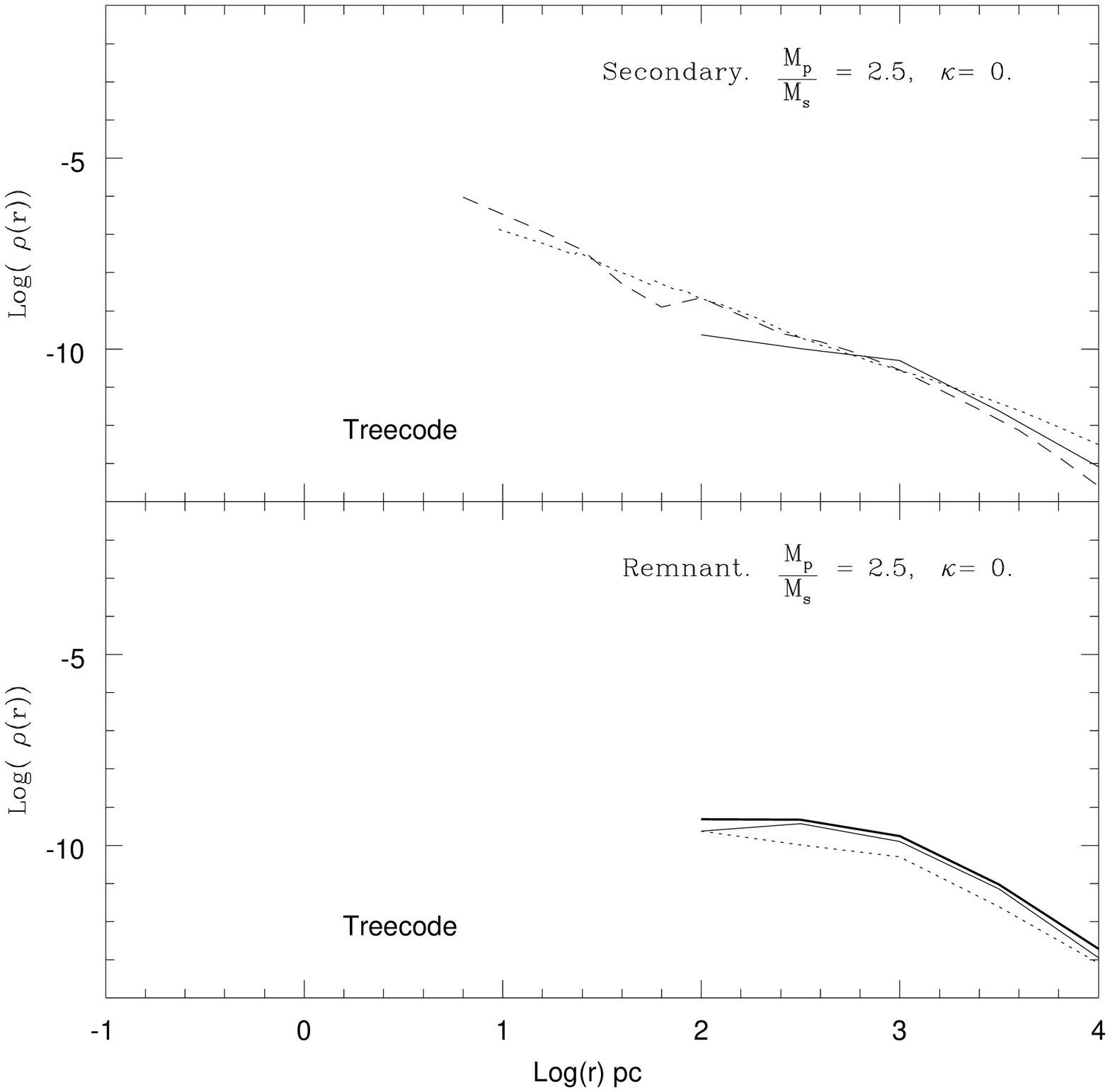]{Treecode comparison for 2.5:1.
See caption for Figure 6, noting that in this experiment, the density profile
of the primary changes. The 
secondary evolves until it is everywhere below the initial state of the primary. Refer to $\S$ 4.5.
\label{Figure 17}}

\figcaption[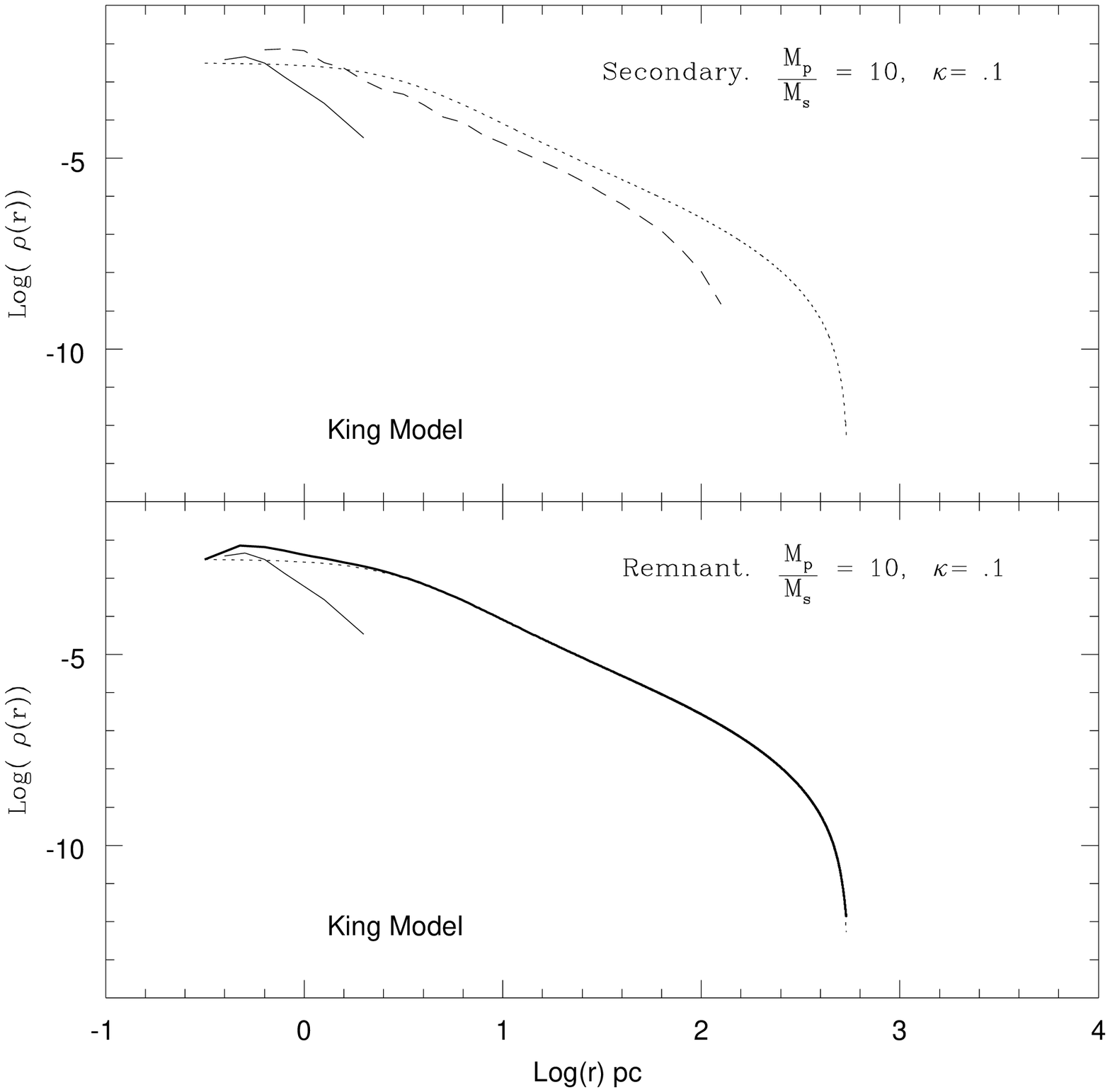]{Comparison to Weinberg's results for 10:1.
See caption for Figure 6.
In this experiment, the secondary is destroyed, as it is in Weinberg's
paper. The secondary central density is only a factor of 3 greater than the primary initially, and after the merger, the density of the secondary is everywhere less than 10 percent greater than the primary. We regard this 
as consistent with Weinberg's experiment.  \label{Figure 18}}

\newpage

% TABLE1.TEX -- AASTeX sample table 1.

% The following gobbledygook is done so that this particular table
% paginates in a way that "looks nice" regardless of whether apjpt4
% or aj_pt4 styles are chosen.  The \tablebreak commands that appear
% after some of the lines force page breaks when the apjpt4 style is
% selected, but act simply as line delimiters for aj_pt4.

\makeatletter
\def\jnl@aj{AJ}
\ifx\revtex@jnl\jnl@aj\let\tablebreak=\nl\fi
\makeatother

% From here on, the file contains tabular data as an author might
% prepare it.

\begin{deluxetable}{lrrrrcrrrrr}
\scriptsize
\tablewidth{0pc}
\tablecaption{Eta Model Galaxy Parameters}
\tablehead{
\colhead{${M_1}\over{M_2}$}     
& \colhead{Galaxy}      
&\colhead{M$_{\rm v}$}          
& \colhead{r$_{\rm core}$\tablenotemark{a} }       
&\colhead{M$_{\rm core}$\tablenotemark{b} }     
& \colhead{$\eta_{ \rm core}$ }    
&\colhead{r$_{\rm env}$ }      
& \colhead{M$_{\rm env}$ }        
&\colhead{$\eta _{\rm env}$ }   
& \colhead{r$_{\rm half}$ }    
&\colhead{t$_{\rm core}$} }
\startdata
100:1 & Primary& -22.0& 263& 4.7 x 10$^{10}$& 2.15 & 4000 &  3.4 x 10$^{12}$ & 2.2 & 10617 & 5932 \nl
&&&&1.34&&&98.66&&\nl
& Secondary& -18.0& 3.8& 1.6 x 10$^{8}$ & 1.0 & 300 & 3.5 x 10$^{10}$ & 1.5 & 508 & 123 \nl
&&&&4.4 x 10$^{-3}$&&&0.996&&\nl
10:1 & Primary& -21.5& 155& 2.3 x 10$^{10}$ & 1.96 & 4000 & 1.9 x 10$^{12}$ &2.0 & 9506 & 9220 \nl
&&&&0.117&&&9.88&&\nl
& Secondary& -19.5 & 18.63 & 1.3 x 10$^{9}$ & 1.23 & 1029 & 1.95 x 10$^{11}$ & 1.5& 1734 & 1113 \nl
&&&&6.7 x 10$^{-3}$ &&& 0.993 &&\nl
2.5:1 & Primary& -21.5& 155& 2.3 x 10$^{10}$ & 1.96 & 4000 & 1.9 x 10$^{12}$ &2.0 & 9506 & 9220 \nl
&&&&2.9 x 10$^{-2}$&&& 2.48 &&\nl
&Secondary & -20.55 & 56.6 & 5.9 x 10$^{9}$ & 1.61 & 2000 & 6.5 x 10$^{11}$ & 2.0 & 4800 & 13101 \nl
&&&&9.3 x 10$^{-4}$ &&& 0.9908 &&
 
\tablenotetext{a}{radii are in units of pc.}
\tablenotetext{b}{the top masses are in units of M$_{\odot}$, and the bottom masses are normalized such that the total secondary mass is 1.0}
\enddata
\end{deluxetable}

\begin{deluxetable}{lrcrr}
\scriptsize
\tablewidth{0pc}
\tablecaption{Synopsis of Experiments}
\tablehead{
\colhead{Type}
& \colhead{Mass Ratio}
& \colhead{$\kappa$}
& \colhead{$f_{\rm drag}$}
& \colhead{Effect on Secondary}}
\startdata
Fiducial & 100:1 & 0.0 & 0.4 & Intact \nl
& 10:1 & 0.0 & 0.1 & Intact \nl
& 2.5:1 & 0.0 & 0.1 & Disrupted \nl
Vary Drag & 100:1 & 0.0 & 0.2 & Intact \nl
& & & 0.1 & Intact \nl
Inner Eta & 100:1 & 0.0 & 0.4 & Intact \nl
& 10:1 & 0.0 & 0.1 & Intact \nl
& 2.5:1 & 0.0 & 0.1 & Disrupted \nl
Non-Plunging Orbits & 100:1 & 0.2 & 0.4 & Intact \nl
& 10:1 & 0.2 & 0.1 & Intact \nl
& & 0.6 & 0.1 & Intact \nl
Treecode & 2.5:1 & 0.0 & N/A & Disrupted \nl
Weinberg & 10:1 & 0.1 & 0.084 & Disrupted 
\enddata
\end{deluxetable}

\begin{deluxetable}{lrrcrrr}
\scriptsize
\tablewidth{0pc}
\tablecaption{King Model Galaxy Parameters}
\tablehead{
\colhead{Mass Ratio}     & \colhead{Galaxy}      &
\colhead{M$_{\rm v}$\tablenotemark{a}}          & \colhead{W$_{0}$ }       &
\colhead{r$_{ \rm core}$\tablenotemark{b}} & \colhead{Mass\tablenotemark{b} }  &
\colhead{$\rho_{0}$\tablenotemark{b} }      }
\startdata
10:1 & Primary& -21.5& 9.5& 3.15 & 10.0& .0032 \nl
& Secondary& -19.5& 9.5& 1.0 & 1.0 & .01
 
\tablenotetext{a}{Magnitude is a free parameter chosen to compare the $\eta$ model results }
\tablenotetext{b}{normalized such that the mass and r$_{\rm core}$ of the secondary is 1.0}
\enddata
\end{deluxetable}

\end{document}